\documentclass[%
 reprint,
 amsmath,amssymb,
 aps,
]{revtex4-1}

\usepackage{graphicx}

\usepackage{subfigure}
\usepackage{dcolumn}
\usepackage{bm}
\usepackage[dvipsnames]{xcolor}
\usepackage{hyperref}
\usepackage{amssymb,amsmath,amsthm}
\usepackage[inline]{enumitem}
\usepackage{tasks}
\usepackage{epstopdf}
\usepackage{color} 
\usepackage{tikz}
\usepackage{calc}  

\tikzset{
    disc/.style={fill=black!60}  
}

\begin{document}

\title{Thermodynamic stability versus Kinetic Accessibility: Pareto Fronts for Programmable Self-Assembly}
\author{Anthony Trubiano and Miranda Holmes-Cerfon}
\affiliation{Courant Institute of Mathematical Sciences, New York University, New York, New York 10012, USA}
\date{\today}

\begin{abstract}
A challenge in designing self-assembling building blocks is to ensure the target state is both thermodynamically stable and kinetically accessible. 
These two objectives are known to be typically in competition, but it is not known how to simultaneously optimize them. 
We consider this problem through the lens of multi-objective optimization theory: 
we develop a genetic algorithm to compute the Pareto fronts characterizing the tradeoff between equilibrium probability and folding rate, for 
a model system of small polymers of colloids with tunable short-ranged interaction energies. 
We use a coarse-grained model for the particles' dynamics that allows us to efficiently search over parameters, for systems small enough to be enumerated.
For most target states there is a tradeoff when the number of types of particles is small, with medium-weak bonds favouring fast folding, and strong bonds favouring high equilibrium probability. The tradeoff disappears when the number of particle types reaches a value $m_*$, that is usually much less than the total number of particles. This general approach of computing Pareto fronts allows one to identify the minimum number of design parameters to avoid a thermodynamic-kinetic tradeoff. However, we argue, by contrasting our coarse-grained model's predictions with those of Brownian dynamics simulations, that particles with short-ranged isotropic interactions should generically have a tradeoff, and avoiding it in larger systems will require orientation-dependent interactions. 
\end{abstract}

\maketitle

\section{Introduction}
Designing novel materials or structures out of mesoscopic building blocks, such as colloids, is a widespread challenge, with applications in drug delivery, optical metamaterials, and environment sensing micro-robots, among others \cite{DrugDelivery1, DrugDelivery2, Optics1, Optics2, Autonomous1, Autonomous2}.  A fundamental issue is to design the building blocks so the desired target state is both \emph{thermodynamically stable}, i.e. it remains near the target state for long times at finite temperature, and \emph{kinetically accessible}, i.e. the system self-assembles into the target state over a reasonable timescale.  
In general, guaranteeing a target state is thermodynamically stable does not guarantee it is kinetically accessible, and vice versa, with these two objectives often working against each other \cite{hagan_viralSA,doye_patchyIcosahedra,whitelam_microscopic_reversibility,palma2012predicting,whitelam_review,Bisker:2018kv}.
Yet, we know from the plethora of successfully self-assembling material and biological systems, that it \emph{is} possible to achieve both objectives simultaneously given the right choice of parameters. Consider proteins: while randomly-designed proteins take astronomically long times to fold into their native states \cite{BRYNGELSON:1995hq,Chan:1994dr,Li:1998tu},
the proteins found in nature fold quickly, a great many orders of magnitude faster than it would take them to sample their thermodynamic equilibrium distribution, a phenomenon known as Levinthal's paradox \cite{levinthal1969fold,levinthal}. 
The reason why is said to be that naturally occurring proteins have evolved so their free energy landscape is ``funnel-shaped'', with the folded structure  at the bottom of the funnel and most dynamic pathways leading downhill, encountering only small free energy barriers along the way \cite{BRYNGELSON:1995hq,Onuchic:1997uq,funnelsDill,Bowman:2010gx}.

Inspired by proteins, can we design material systems to similarly fold quickly to a deep metastable state? 
That is, given a collection of building blocks with various design parameters (such as interaction energies, concentrations, shapes, etc), can we choose parameter values so the system achieves both objectives of thermodynamic stability and kinetic accessibility simultaneously? 
This question lies at the heart of most theoretical studies of programmable self-assembly, but it is a challenge to optimize two objectives simultaneously. 
To simplify, earlier studies of self-assembly considered evaluating measures of thermodynamic stability and kinetic accessibility as design parameters are varied along an axis; such studies found there is a narrow range of parameter values in the observation regions where assembly may occur, and furthermore that 
the finite time yield of the target state is non-monotonic in the binding energy \cite{hagan_viralSA, whitelam_review, whitelam_microscopic_reversibility}. 
More recent studies have considered self-assembly in an optimization framework, 
by building methods to optimize a single objective measuring progress toward the target state. One such objective is the partition function, measuring thermodynamic stability; 
 much progress has been made in understanding how to choose pair interaction energies between distinct particles to make a target structure low free energy \cite{brennerDesign, zoranaSizeLimits,whitelam_growth,miskinEngine,Madge:2017gc,Das:2021jk}, motivated partly by the spectacular self-assembly of DNA-brick based structures \cite{Ke:2012jdd}. Adjusting the concentrations of the components of a target structure can further lower its free energy
\cite{undesiredUsage}. 
Kinetic effects, while harder to study, 
have been 
controlled via nucleation barriers \cite{Jacobs:2015gia}, geometrical measures of the ruggedness of the free energy landscape \cite{funnelJinWang}, or by maximizing the yield of a target structure at a fixed time\cite{LongML,whitelamML}, an approach which couples thermodynamic and kinetic considerations into a single objective. Other measures of kinetics may be tuned by building automatic differentiation into molecular dynamics simulations, which promises to be a powerful tool for optimizing design parameters of a self-assembling system \cite{Goodrich:2021fa}.

In this paper we take an alternative approach: 
rather than optimizing for a single measure of the success of programmable self-assembly, we consider the two objectives of thermodynamic stability and kinetic stability separately, and seek to understand and ultimately to minimize the tradeoff between these two objectives as a the design parameters are varied. 
%
We focus our examples on a model system, namely polymers of colloids interacting with short-ranged isotropic attractive forces, where the design parameters are the energies of the attractive interactions. Such particles can be synthesized experimentally by gluing single-stranded DNA onto the surfaces of the colloids, and are a promising set of building blocks for designing new materials \cite{Macfarlane:2011fh,dna1, dna2, Wang:2017bd, McMullen:2018iy}. We consider small collections of colloids, which have been thoroughly studied both experimentally \cite{mengExp,experimentColloidDiffusion} and theoretically \cite{zoranaSizeLimits, brennerDesign, zoranaLivingMatter}. We coarse-grain their dynamics to efficiently search over parameters, which allows us to 
 exhaustively characterize their thermodynamic-kinetic tradeoffs for this coarse-grained system, a pedagogical characterization that has not been illustrated before. 

We characterize the thermodynamic-kinetic tradeoff for a target structure by computing its \textit{Pareto front}, namely the set of values attained by the objectives such that it is not possible to further increase both objectives \cite{Stewart2008}. 
The Pareto front is part of the boundary of everything that is achievable by the objective functions as the design parameters are varied.  
This curve gives qualitative insight into what sets the tradeoff between thermodynamics and kinetics for different target structures, and it is useful as a practical design tool: one should always choose parameter values that give objective values on this curve, since otherwise one could increase both objectives using different parameters. 
Furthermore, the Pareto front allows a user to \emph{control} the tradeoff between the objectives, by deciding how many design parameters to vary. As the number of design parameters increases, the Pareto front necessarily becomes steeper, implying the tradeoff between the objectives becomes less significant. Strikingly, in our examples, there is a sharp transition where the tradeoff suddenly becomes minimal. We argue that this is a signature of a ``funnel-shaped'' landscape. 

The transition to a steep Pareto front gives the minimal number of design parameters to avoid a tradeoff, within the context of our coarse-grained dynamical model. Sometimes the same design parameters also lead to fast folding in Brownian dynamics simulations, and sometimes they don't. When they don't, we show it is because of ``chiral traps'', configurations that cannot reach the target state without breaking a strong bond, but that have the same bonds as a cluster on the correct folding pathway. We argue that because of these inevitable chiral traps, particles with short-ranged isotropic interactions should generically have a tradeoff between thermodynamics and kinetics; avoiding this tradeoff in larger systems will require particles with orientation-dependent interactions.

\section{There is a tradeoff between thermodynamic stability and kinetic accessibility}

\begin{figure}
    \centering
    \def\sc{0.45}
    \begin{tabular}{cccc}
    \begin{tikzpicture}[thick,scale=\sc]
\filldraw[disc] (0,0) circle (0.5);
\filldraw[disc] (1,0) circle (0.5);
\filldraw[disc] (0.5,0.866) circle (0.5);
\filldraw[disc] (1.5,0.866) circle (0.5);
\filldraw[disc] (1,1.7321) circle (0.5);
\filldraw[disc] (0,1.7321) circle (0.5);
\filldraw[disc] (-0.5,0.866) circle (0.5);
\end{tikzpicture}
&
\begin{tikzpicture}[thick,scale=\sc]
\filldraw[disc] (0,0) circle (0.5);
\filldraw[disc] (1,0) circle (0.5);
\filldraw[disc] (0.5,0.866) circle (0.5);
\filldraw[disc] (1.5,0.866) circle (0.5);
\filldraw[disc] (2.5,0.866) circle (0.5);
\filldraw[disc] (1,1.7321) circle (0.5);
\filldraw[disc] (2,1.7321) circle (0.5);
\end{tikzpicture}
&
\begin{tikzpicture}[thick,scale=\sc]
\filldraw[disc] (0,0) circle (0.5);
\filldraw[disc] (1,0) circle (0.5);
\filldraw[disc] (2,0) circle (0.5);
\filldraw[disc] (0.5,0.866) circle (0.5);
\filldraw[disc] (1.5,0.866) circle (0.5);
\filldraw[disc] (2.5,0.866) circle (0.5);
\filldraw[disc] (3,0) circle (0.5);
\end{tikzpicture}
&
\begin{tikzpicture}[thick,scale=\sc]
\filldraw[disc] (0,0) circle (0.5);
\filldraw[disc] (1,0) circle (0.5);
\filldraw[disc] (2,0) circle (0.5);
\filldraw[disc] (0.5,0.866) circle (0.5);
\filldraw[disc] (1.5,0.866) circle (0.5);
\filldraw[disc] (1,1.7321) circle (0.5);
\filldraw[disc] (0.5,-0.866) circle (0.5);
\end{tikzpicture}
\\
 12  bonds & 11 bonds & 11 bonds & 11 bonds
    \end{tabular}
    \caption{
    A system of 7 identical disks that illustrates the tradeoff between thermodynamic stability and kinetic accessibility. The leftmost, ``flower'' cluster, has the highest equilibrium probability  when the interactions are strong, because it has the most bonds. However, the average time to form the flower increases exponentially with the energy of the interaction, because most trajectories hit another clusters first \cite{klein2019structure,folding}. 
    }
    \label{fig:flowertradeoff}
\end{figure}
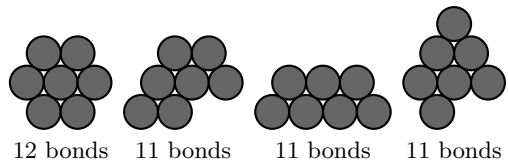

\begin{figure*}[t]
\centering  
\includegraphics[width=\linewidth]{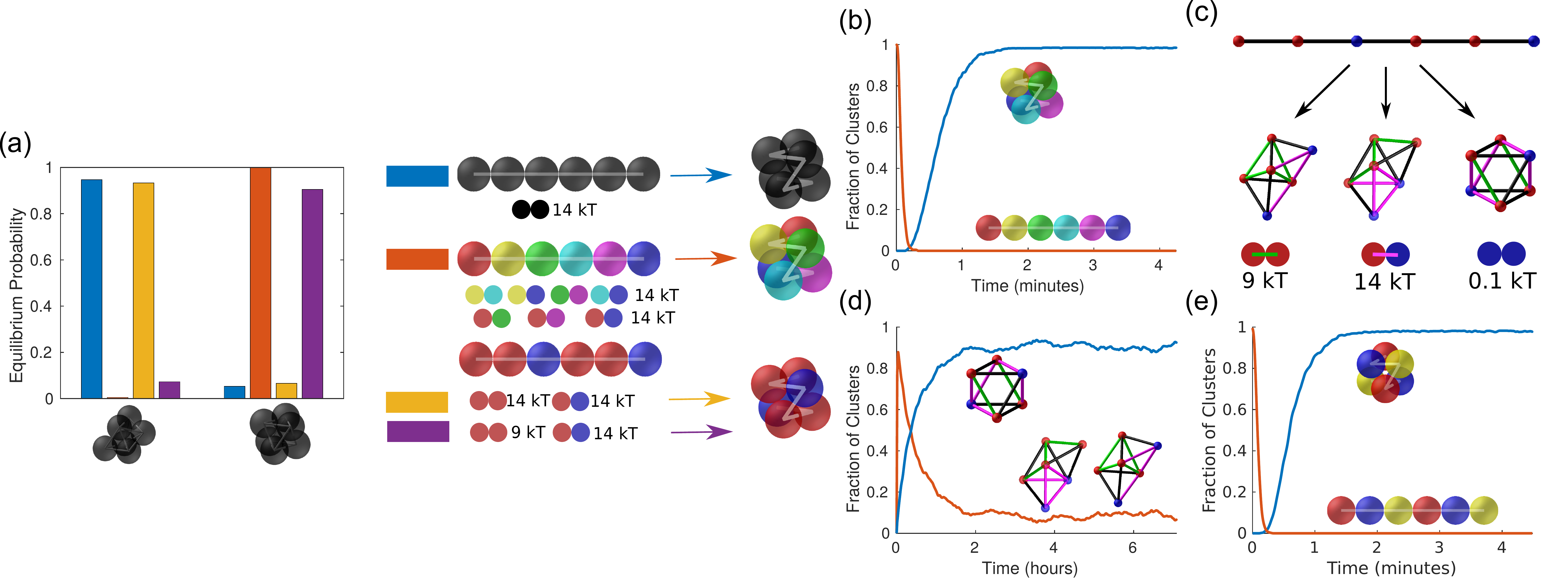}
\caption{There is a thermodynamic-kinetic tradeoff to form an octahedron of 6 spheres, but this can be avoided with the right choice of particle types and parameters. 
(a) Equilibrium probabilities for low energy clusters of polymers of 6 spheres, the polytetrahedron (left) and octahedron (right), with different design parameters (bond energies). The bond energies are listed in the adjacent legend, for: identical particles, distinct particles, two types of particles arranged in an $AABAAB$ configuration (A=red, B=blue).
Any unlabeled interactions are $0.1 kT$. 
(b) The octahedron forms within \emph{minutes} from a linear chain in a Brownian dynamics simulation, when formed from distinct particles with interactions given above. The timescale is estimated for 1.3 $\mu$m colloids \cite{experimentColloidDiffusion}. 
(c) Competing ground states for the AABAAB chain, showing AA bonds (green), AB bonds (pink),  and backbone bonds (black). 
(d) The octahedron forms within \emph{hours} when made from two types of particles  (interactions given in purple legend), because of the presence of kinetic traps. 
(e) With three types of particles and optimal interactions $E_{AB}{=}E_{AC}{=}E_{BC}{=}14 kT$ (all other pairs  $0.1 kT$),  the octahedron forms within minutes, as for distinct particles. There is no more thermodynamic-kinetic tradeoff. 
}\label{fig:n6}
\end{figure*}


A simple example that illustrates the tradeoff between thermodynamic stability and kinetic accessibility for particles with short-ranged attractive interactions is shown in Figure \ref{fig:flowertradeoff} \cite{klein2019structure}. This figure shows  clusters of 7  disks in the plane, that are are energetic local minima when the particles interact with a short-ranged, isotropic pairwise attractive potential.
For identical disks, when the bond energy $E$ is large, then the leftmost, ``flower'' cluster has the lowest free energy, because it has 12 bonds whereas the other clusters have only 11 bonds. Therefore, as $E$ increases, the equilibrium probability of the flower (when sufficiently confined) approaches 1, making it thermodynamically stable. However, as $E$ increases, the average time it takes to form the flower increases too, because most dynamical pathways hit one of 11-bond clusters first \cite{klein2019structure,folding}, 
and breaking a bond to reach the flower happens on a timescale proportional to $e^{E/kT}$. Therefore, the flower becomes less kinetically accessible as it becomes more thermodynamically stable.

This tradeoff occurs even when the particles are non-identical. Consider now a system of $6$ spheres which are constrained to stay in a chain, as a polymer. We add this backbone constraint, partly to mimic proteins, and partly to obtain a system that does not evaporate without introducing undesired features like walls. 
We assume here and throughout the paper that particles interact with a pairwise Morse potential, $U(r) = E\left(e^{-2\rho(r-d)}-2e^{\rho(r-d)}\right)$, with diameter $d{=}1$, range parameter $\rho{=}40$, and an energy $E$ of the interaction that varies depending on the particle pair. 
There are two clusters that are energetic local minima when $E$ is the same for all pairs, 
an octahedron and a polytetrahedron (Figure \ref{fig:n6}(a)). Both have 12 bonds hence the same potential energy, but their differing rotational and vibrational entropies imply that when $E{=}12kT$, the octahedron has an equilibrium probability  of only $0.0535$ -- compare this to the polytetrahedron's equilibrium probability of $0.9456$ (SI Appendix$^\dag$, Section \ref{sec:EqProb}).  The remaining probability is associated with floppy clusters.

Let's try to make the octahedron have high probability, by letting the particles be distinct.  
Following the principles of \cite{brennerDesign, zoranaLivingMatter, whitelam_growth}, 
which claim that optimal assembly occurs when 
bonds in the target state are identically strong and other bonds are identically weak, 
choose $E{=}14 kT$ for interactions in the target octahedron, while $E{=}0.1 kT$ for all other interactions. The weak bonds have non-zero energy, to avoid computational difficulties with exact zeros, and because it is hard to make DNA-coated colloids that have strictly zero attraction \cite{Huntley:2016dn}. With these interactions, the octahedron has equilibrium probability $0.99995$ (Figure \ref{fig:n6}(a)). 
It furthermore assembles quickly, with few kinetic traps: Figure \ref{fig:n6}(b) shows that in Brownian dynamics simulations starting from a linear chain, the octahedron reaches a nearly steady probability of $0.985$ after about 1 minute, where the timescale is estimated for micron-scale colloids using the experimental parameters in \cite{experimentColloidDiffusion} (SI Appendix$^\dag$, Section \ref{sec:BD}). The yield is not perfect because 
the chain sometimes reaches a floppy state in which existing bonds geometrically block additional correct bonds from forming. 
We return to this issue in Section \ref{sec:chiral}, but for now use these results for six particle types as a benchmark.


Making all particles distinct with controlled interactions is currently not feasible for DNA-coated colloids, and furthermore, biology tells us we should not require a number of particle types proportional to system size -- proteins form tens of thousands of distinct structures from strings of hundreds to thousands of only 20 types of amino acids. To see if we can make the octahedron from fewer types of particles, we consider a chain with only two particles types, A and B, ordered along the chain as AABAAB. Then the target configuration has two AA bonds and five AB bonds.  Naively applying the 
same principle as before, 
we set $E_{AA} {=} E_{AB} {=} 14 kT$ and $E_{BB} {=} 0.1 kT$. 
The octahedron now has an equilibrium probability of $0.08$. This is still small, because there are two polytetrahedra with the same total number of AA and AB bonds, hence the same potential energy, but which are favoured entropically (Figure \ref{fig:n6}(c)). 

As might have been expected, making target bonds strong and non-target bonds weak is not sufficient to form a target state when the number of particle types is restricted.  However, a bond-counting argument suggests the octahedron can form if we additionally allow bonds with \emph{medium} strength. The octahedron has five $AB$ bonds while the competing states only have three or four, and it has two $AA$ bonds while the competing states have four or three (Figure \ref{fig:n6}(c)). By making $AB$ bonds strong and $AA$ bonds have medium strength, the potential energy of the octahedron should be lowered compared to the polytetrahedra. 
%
With $E_{AB}{=}14 kT$, $E_{AA}{=}9 kT$,
the equilibrium probability for the octahedron is now $0.905$ (Figure \ref{fig:n6}(a)). As  $E_{AA},E_{AB} \to \infty$ with $E_{AB}-E_{AA} \to \infty$, the equilibrium probability approaches $1$.

Having identified bond energies that give the octahedron high equilibrium probability, we  test whether it forms in a dynamical simulation.
Figure \ref{fig:n6}(d) shows the octahedron reaches a nearly steady probability of about $0.9$, but only after about 2 hours.
Compared to the  timescale of 1 minute for distinct particles, this is significantly slower, more than two orders of magnitude slower.



The octahedron forms  slowly because the polytetrahedra are kinetic traps: as the octahedron's free energy decreases (by increasing $E_{AA},E_{AB}$), so too does the polytetrahedron's free energy, and simultaneously the energy barrier to transition from a polytetrahedron to the octahedron increases. Many trajectories hit a polytetrahedron before the octahedron, and must break AA bonds to form the octahedron, which happens slowly. 
With two types of particles arranged as AABAAB, it appears there is a tradeoff between the thermodynamic stability and the kinetic accessibility of the octahedron.


\section{Quantifying the tradeoff between thermodynamic stability and kinetic accessibility}

\begin{figure}[h]
\centering  
\includegraphics[width=1.0\linewidth]{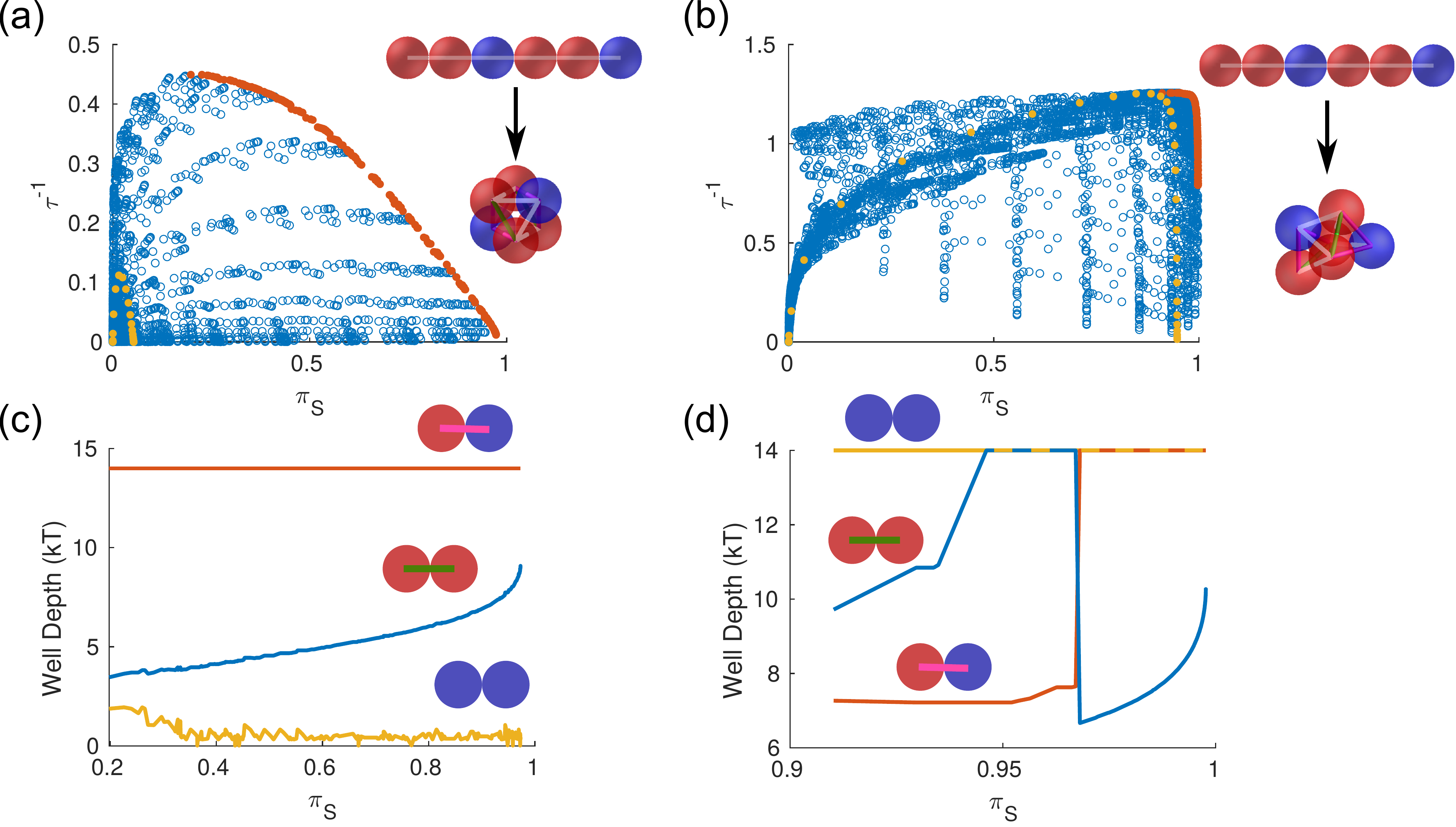}
\caption{
Quantifying the tradeoff between equilibrium probability $\pi_S$ and folding rate $\tau^{-1}$ for a chain of AABAAB spheres. 
(a,b) Scatter plots ($\pi_S,\tau^{-1}$)-values (outcomes) to each of the two rigid clusters, (a) octahedron and (b) polytetrahedron. The octahedron forms the permutation shown plus one other, while the polytetrahedron forms $12$ energetically equivalent permutations which are lumped together. Blue points are outcomes evaluated with bond energies $E_{AA},E_{AB},E_{BB}$ on a grid, each taking $20$ equally spaced values in $[0.1,14]$. 
Yellow points are outcomes with $E_{AA}{=}E_{AB}{=}E_{BB}$, hence correspond to identical particles. 
Orange points, on the boundary of the set of outcomes, form the Pareto front, computed using our genetic algorithm. 
(c,d) Bond energies along the Pareto fronts for each cluster, as a function of the equilibrium probability $\pi_S$.  
}\label{fig:paretoAABAAB}
\end{figure}

Can we choose bond energies such that the octahedron forms quickly yet has high equilibrium probability? And can we further understand the tradeoff, for this and other systems? 
We answer these questions by framing them as a multi-objective optimization problem: we wish to simultaneously maximize two functions, one measuring thermodynamic stability and one measuring kinetic accessibility. Usually it is not possible to maximize both functions at once, in which case we wish to quantity the tradeoff between the two. 

To proceed, let us measure thermodynamic stability by $\pi_S$, the equilibrium probability of all configurations $S$ that correspond to a target state, such as the octahedron, and let us measure kinetic accessibility by $\tau^{-1}$, a function we call the \emph{rate}, where $\tau$ is the mean first passage time (mfpt) from the linear chain to the target state. The functions $\pi_S$ and $\tau^{-1}$ are our \emph{objective functions}, or \emph{objectives}, the functions we wish to maximize. We have a collection of design parameters, such as the  interaction energies between particles of different types. The values of the objectives change as the design parameters vary. 

A technical but important issue is evaluating the objectives for different design parameters. Evaluating these by brute-force simulation is extremely time-consuming, so we instead construct a coarse-grained model that lets us efficiently evaluate the objectives for different design parameters. Our model starts by lumping together configurations that share the same adjacency matrix,  and then approximates the dynamics as a Markov chain on the set of  feasible adjacency matrices, including those corresponding to floppy configurations. The rates of forming a bond are estimated from simulations and are assumed to be independent of bond energies, and the rates of breaking a bond are set by detailed balance, where the equilibrium probabilities for each node of the Markov chain (each adjacency matrix) are estimated by a Monte Carlo sampling procedure with fixed bond energies \cite{HolmesCerfon:2020dk} (SI Appendix$^\dag$, Section \ref{sec:coarse}). Obtaining the initial rates and equilibrium probabilities with one set of bond energies is time-consuming, but once these are calculated we may reweight them for other energies. Then, calculating $\pi_S$ and $\tau^{-1}$  simply requires solving a linear algebra problem for each different set of bond energies. This approach is only feasible for systems small enough that all the adjacency matrices (including permutations) are possible to enumerate.

Let's explore our model for different types of particles. 
With one type of particle we may vary the energy $E$ of the common interaction. Figure \ref{fig:paretoAABAAB}(a) (yellow dots) shows 
the rate $\tau^{-1}$ is maximized for medium-weak energies, $E\approx 5.2$. At this energy $\pi_S$ is small because the floppy configurations are also probable. As $E$ increases beyond this value, $\tau^{-1}$ decreases and $\pi_S$ increases, approaching their limiting values of $(\pi_S,\tau^{-1}) = (0.054,0)$ as $E\nearrow \infty$. The set of achievable $(\pi_S,\tau^{-1})$-values, called the set of \emph{outcomes}, describes the tradeoff between stability and accessibility for identical particles.

Now  consider a chain of two types of particles arranged along a chain as AABAAB. 
We vary the three energies $\vec E = (E_{AA},E_{AB},E_{BB})$ on a grid. Figure \ref{fig:paretoAABAAB}(a) (blue dots) shows a scatter plot of the outcomes $(\pi_S, \tau^{-1})$. These outcomes appear to fill a two-dimensional region of $(\pi_S, \tau^{-1})$-space. This region has expanded well beyond the outcomes for identical particles, with a maximum rate about four times higher and a maximum equilibrium probability approaching $1$. Yet, there are still no outcomes that maximize $\pi_S$ and $\tau^{-1}$ simultaneously. 
As for identical particles, the parameters that give the maximum rate, $\vec E = (3.5, 14, 1.9)$
are not all large -- fast folding appears to require some weak bonds. 
Also as for identical particles, there are parameters such that $\pi_S\to 1$, but in this limit $\tau^{-1}\to 0$.

A key observation is that the set of outcomes appears to be bounded -- in between the outcomes with maximum rate, and maximum equilibrium probability, the outcomes appear to  lie on one side of a curve. 
This curve, which we have approximated using an algorithm to be described later (orange points), will play a key role in our analysis. 
We argue that if one wishes to design a self-assembling system, one should choose parameters that give outcomes on this curve. 
To see why, suppose one chooses parameters that give an outcome away from the curve,  in the interior of the set of outcomes. Then there exist outcomes, to the right and upward, that have higher equilibrium probability (without decreasing the rate), higher rate (without decreasing the equilibrium probability), or both. One can continue to vary parameters to improve either objective until one reaches the boundary of the set of outcomes, where one can no longer increase both objectives simultaneously.

In the language of multi-objective optimization, an outcome such that no objective value can be increased without decreasing the value of another is called \emph{Pareto optimal} \cite{Stewart2008}. The collection of all such outcomes is called the \emph{Pareto front}. The curve that bounds the set of outcomes in Figure \ref{fig:paretoAABAAB}(a) is a Pareto front. 

The Pareto front is a powerful design tool. 
With this curve in hand, a designer can choose an outcome that weights the objectives according to their interests, and then read off the parameter values (bond energies) that give this outcome. 
For example, a designer who wishes to obtain at least 50\% equilibrium yield, might choose parameters that give the maximum $\tau$ subject to $\pi_S>0.5$ so the system also assembles as quickly as possible; a designer with 
a limited time budget might choose parameters that give a value of $\tau$  comparable to their desired timescale, ensuring the system is likely to reach the target state but also has the best probability of remaining at it. 
This approach of providing a set of optimal values, which is common in design problems that consider multiple objectives (e.g. building a product that has high quality but is not too costly, creating a cancer radiation treatment plan that is effective but safe, designing airfoils that are energy efficient while minimizing sound pressure) \cite{Stewart2008}, is agnostic to which objective function is better than the other; it allows different designers to place different levels of importance on the objectives. 


\section{Pareto fronts for a collection of examples} \label{paretoGA}

\begin{figure}[h]
\centering  
\includegraphics[width=1.0\linewidth]{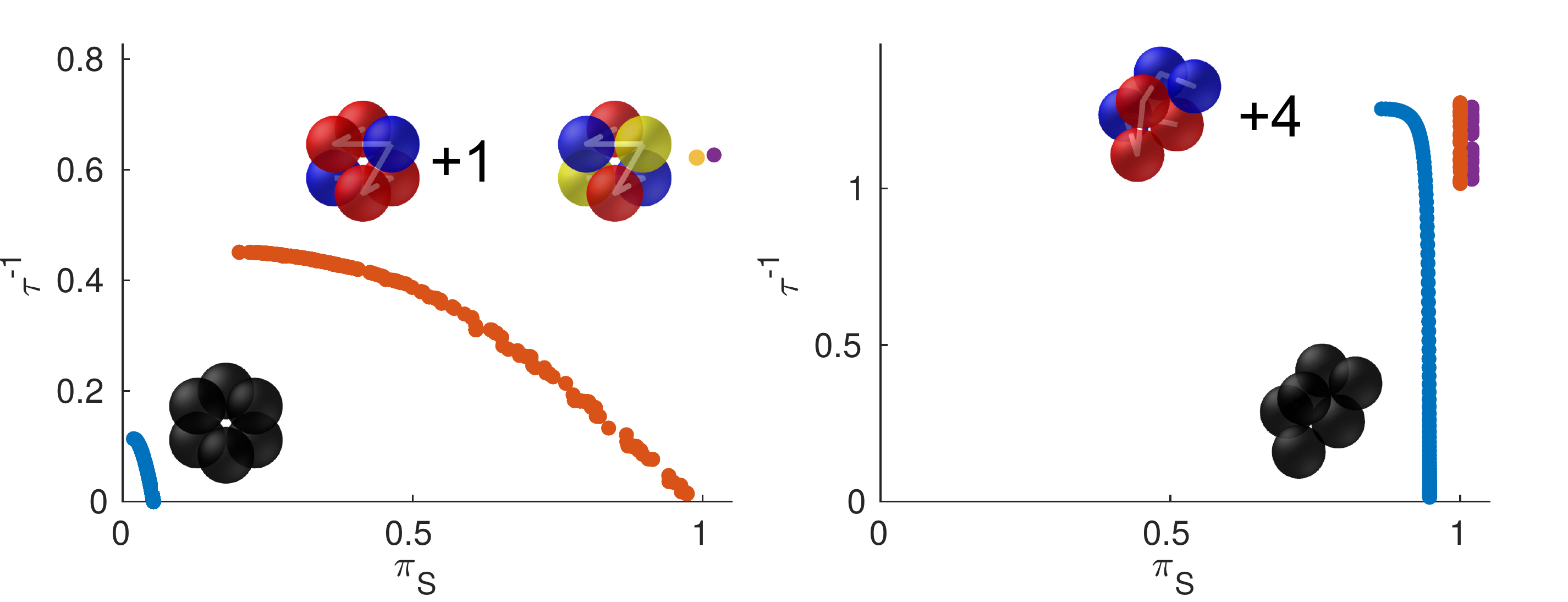}
\caption{
Pareto fronts for polymers of 6 spheres with $m$ particle types: $m{=}1$ (blue), $m{=}2$ (orange), $m{=}3$ (yellow), $m{=}6$ (purple), for the octahedron (left) and polytetrahedron (right). 
The octahedron's Pareto front with $m{=}3$ types is a single point, implying there is no thermodynamic-kinetic tradeoff. The polytetrahedron's Pareto front with $m{=}2$ types is essentially vertical, implying the same. 
The optimal particle labels are AABAAB and ABCABC for the octahedron, and ABAABB for the polytetrahedron.  
}
\label{fig:paretoSpheres}
\end{figure}

Since the Pareto front describes the fundamental tradeoff between thermodynamic stability and kinetic accessibility as the bond energies are varied,  we turn our attention to computing this curve for systems described by our coarse-grained model. 
We compute the Pareto front using a genetic algorithm. 

The genetic algorithm works as follows. 
We initialize a population of $P$ members with bond energies sampled randomly from a uniform distribution on  $[0.1, E_M]$, where $E_M=12{-}14$ depending on the system. 
At each step of the algorithm, we evaluate the objectives $\pi_S,\tau^{-1}$ for each member and then use these outcomes to compare members. We say member X is \emph{dominated by} member Y if member Y has larger values in each objective than member X. We sort the members by the number of dominating members, from least to greatest, and use this list to construct the next generation.
All non-dominated members, those dominated by no other members, are carried to the next generation. The rest of the population dies, and is replaced by offspring. The offspring are created by choosing two random parents, from the top $p\%$ of the current generation, and choosing the child's parameters by randomly assigning them from each parent. There is also an $r\%$ chance the child undergoes mutation for any given parameter, meaning that parameter is re-sampled from the original distribution. We used $r=10$ and $p$ in the range $[30,50]$. 
We repeat this procedure for each generation until one of two end conditions is met: either the iteration limit is reached, or all members are non-dominated (Appendix, Section \ref{sec:GAdetails}). 

This algorithm also allows us to consider different particle labelings,  e.g. AABBAA, ABBBBA, BAAAAA, etc, by allowing particle type to be an inheritable parameter in our genetic algorithm, assuming a fixed number $m$ of types.
%
During mating, the type of particle $i$ in the child configuration is chosen randomly from the two parents. 
If it mutates, the types is sampled uniformly from the collection of $m$ types. This adds a negligible amount of computation per iteration, at the cost of increasing the number of iterations until convergence. 


\subsection{Octahedron and polytetrahedron} 

The Pareto front for the AABAAB octahedron is shown in Figure \ref{fig:paretoAABAAB}(a) (orange points). This curve lies on the boundary of the set of outcomes, verifying the algorithm is working correctly. From the members on the Pareto front one obtains a parameterization of the bond energies as a function of either objective. Figure \ref{fig:paretoAABAAB}(c) shows the bond energies as a function of $\pi_S$. The strong bond $E_{AB}$ is constant at the maximum value, 
the weak bond $E_{BB}$ fluctuates near the minimum value,  
and $E_{AA}$ increases smoothly along the Pareto front, from a medium-small value where the rate $\tau^{-1}$ is highest, to a larger value where the equilibrium probability $\pi_S$ is highest.

We repeated this calculation for the AABAAB polytetrahedron (Figure \ref{fig:paretoAABAAB}(b)).  
The Pareto front is a union of two curves, one nearly horizontal and one nearly vertical. 
Each curve favors a different permutation of the polytetrahedron with different numbers of $AA$, $AB$ bonds. 
The Pareto front does not extend down to $\tau^{-1}\rightarrow 0$ (when $E_{BB}$ is bounded) because as $E_{AA}$ increases, the equilibrium probability begins to decrease as the particles become identical. 
A notable difference from the octahedron is that the polytetrahedron's Pareto front is much steeper. This means the tradeoff between thermodynamic stability and kinetic accessibility is smaller, since a steeper slope means the system can still achieve a high rate without sacrificing much equilibrium probability. 

\begin{figure*}[ht!]
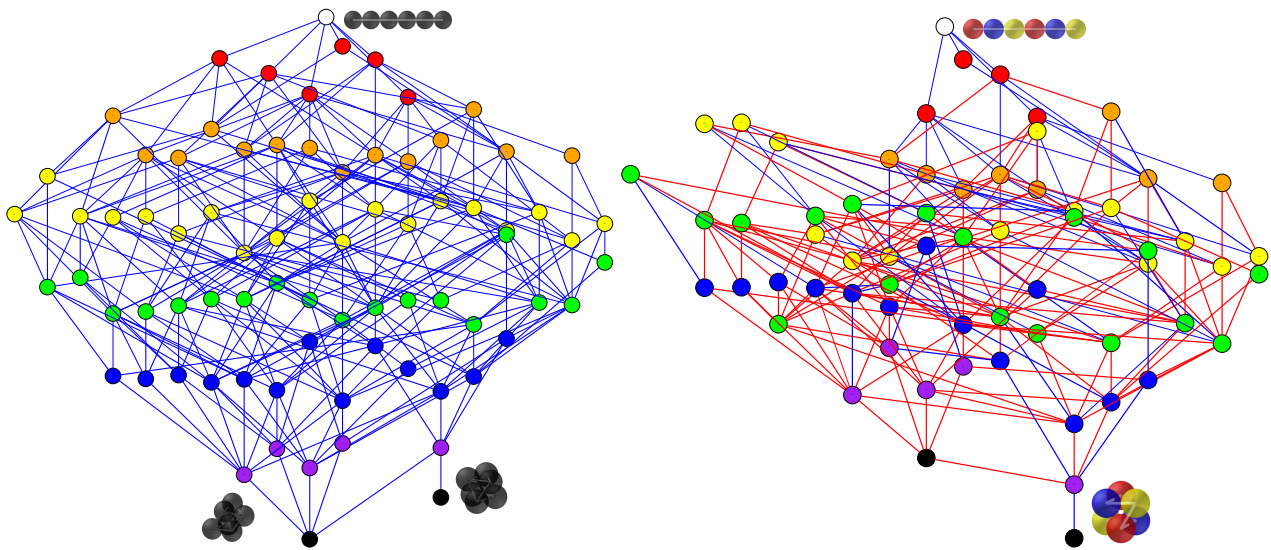

\centering  
\includegraphics[width=0.45\linewidth]{final_figures/fe_identical.pdf}
\includegraphics[width=0.48\linewidth]{final_figures/fe_funnel.pdf}
\caption{Free energy diagrams for the self-assembly of $6$ spheres with identical particles with interaction energy $8 kT$ (left) and three particle types with Pareto optimal interactions for an octahedron (right). The optimal interactions lead to a diagram that appears qualitatively to be ``funnelled.'' 
The vertical positions of the nodes are proportional to the state's free energy. Node coloring denotes the number of non-trivial bonds; the chain in white, one additional bond for each color of the rainbow, and black for ground states. Edge width is proportional to the absolute net transition rate, $|Q_{ij}-Q_{ji}|$, where $Q$ is the transition matrix. Edge color denotes the preferred transition direction: blue for forming a bond, and red for breaking a bond. 
}
\label{fig:free_energy}
\end{figure*}

How can we design a system so its Pareto front is steep? So far we have fixed the particle labels, but allowing them to vary gives greater design flexibility. 
Figure \ref{fig:paretoSpheres} shows the Pareto fronts for the octahedron and polytetrahedron when particle label is included as a design parameter, for various numbers of types $m$.  With 2 types, the octahedron's optimal labeling is the AABAAB that we have been considering and its Pareto front has a shallow slope. With 3 types, however, the Pareto front is  a \emph{single point} -- equilibrium probability and rate are maximized simultaneously. There is no thermodynamic-kinetic tradeoff -- we have found a system with a funnelled energy landscape. To make this analogy with funnelled landscapes more explicit, Figure \ref{fig:free_energy} shows a free energy diagram over the coarse-grained states. We see that for identical particles, the landscape is entirely downhill, with most probability flowing into the global free-energy minimum, the polytetrahedron. For three particle types with the Pareto optimal parameters, all probability flows into the octahedron, which now lies at the bottom of the funnel, while probability only flows out of the polytetrahedron; it is no longer a kinetic trap. 

To check if this optimization works outside the assumptions of our coarse-grained model, we performed Brownian dynamics simulations of chains of 6 spheres using the optimal ordering ABCABC and optimal bond energies found by the algorithm. 
Figure \ref{fig:n6}(e) shows the octahedron forms with high probability in just over 1 minute. 
This result is comparable to our earlier Brownian dynamics simulation with distinct particles, showing that we can achieve the same assembly efficiency using three types of particles as with six types.

\subsection{Polymers of disks}

\begin{figure*}[ht!]
\centering  
\includegraphics[width=1.0\textwidth]{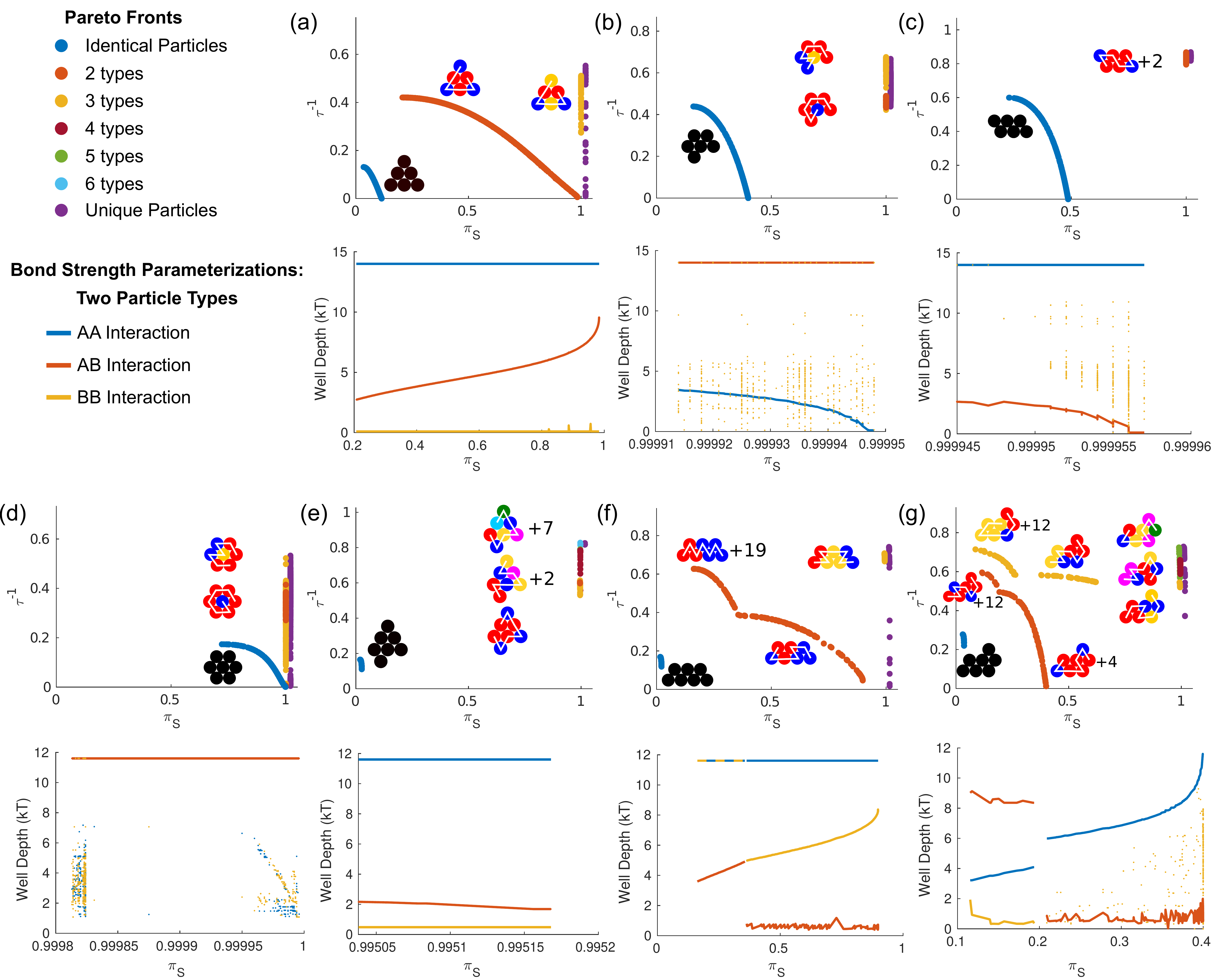}
\caption{Pareto fronts for all rigid clusters of 6 and 7 disks, with varying number of particle types $m$. The target states include all permutations of a given cluster. 
For each target states we calculate the Pareto front for distinct particles (purple) to determine an upper bound on $\tau^{-1}$. 
Then, we increase $m$ until the assembly is as efficient as having distinct particles. Pareto fronts for distinct particles are shifted rightwards so they do not overlap other Pareto fronts. 
Cluster plots show which permutation generate the Pareto front. In some cases, several permutations with approximately equal probability generate the front, indicated by `$+c$', where $c+1$ is the total number of permutations. Below each Pareto front there is a parameterization of the $2$-type Pareto front in terms of the three bond energies. Dotted data indicates the Pareto front is insensitive to the given interaction, for example a $BB$ interaction in a system with only one $B$-type particle.
}
\label{fig:paretoCollection}
\end{figure*}

Figure \ref{fig:paretoCollection} shows the Pareto fronts for all rigid clusters formed from chains of $6$ \& $7$ discs, optimized over particle labels for a variety of types $m$. There are many things to observe about these curves.
\medskip

\emph{They all have a critical value $m{=}m_*$ for which the Pareto front becomes essentially vertical.} This is perhaps our most striking observation. In Figure \ref{fig:paretoCollection}, $m_*{=}2$ for (b,c,d,e) and $m_*{=}3$ for (a,f,g).
 Although the Pareto front does not always collapse to a single point, as it did for the octahedron, when it is very steep
 the tradeoff between thermodynamic stability and kinetic accessibility has been eliminated: a target state may have equilibrium probability close to $1$, and still fold rapidly. 
We suggest that a steep Pareto front is a signature of a funnelled energy landscape. 

\emph{Some clusters can fold faster when the number of types is increased beyond $m_*$}
For example, it appears that the chevron can fold slightly faster with $m{=}3$, even though it has $m_*{=}2$. We confirmed this observation by Brownian dynamics simulation, where we observe a $20\%$ reduction in the mean first passage time by using $m{=}3$ instead of $m{=}2$ (See Figure \ref{fig:chevSims} in the SI Appendix$^\dag$).

\emph{Pareto fronts can be discontinuous.} For example, see Figure \ref{fig:paretoCollection}(g), $m{=}2,3$. Each continuous fragment of the Pareto front is associated with different optimal particle labels. 

\emph{Shallow-sloped Pareto fronts have medium-weak bonds where folding is fastest, and strong bonds where free energy is lowest.} For $m=2$ types, shallow Pareto fronts occur in Figure \ref{fig:paretoCollection}(a,f,g) and for the octahedron, Figure \ref{fig:paretoSpheres}(a). 
For all of these, as $\pi_S$ increases, the strongest bond remains constant  while a weaker bond increases in strength, from a small (but not the smallest) value, to close to the maximum value. 
This case bears comparing with proteins, which are known to be marginally stable, i.e. with energies just low enough to be stable but not as low as could be possible \cite{Taverna:2002gj}. The reasons for proteins' marginal stability are not fully known; one speculation is that it aids proteins' functionality, by giving them some flexibility and also allowing them to be easily taken apart; another is that it is a result of neutral evolution, which favours native structures that can be folded by the most amino acid sequences \cite{Taverna:2002gj}. Our results suggest that weaker bonds could also help proteins fold more quickly, by minimizing the effect of kinetic traps.

\emph{Steeply-sloped Pareto fronts favor (slightly) weaker bonds where free energy is lowest.} 
For $m{=}2$ types, steep Pareto fronts occur in Figure \ref{fig:paretoCollection}(b,c,d,e) and for the polytetrahedron, Figure \ref{fig:paretoSpheres}(b). 
For most of these examples, the strong bond is still constant but the weaker bond decreases slightly as $\pi_S$ increases. 
We suggest three possible explanations. The first is an effect of permutation lumping. When computing the mfpt to a target, all permutations are considered, even if some have vanishingly small equilibrium probability due to the presence of a weak bond. Since the mfpt calculation only considers hitting a target state, and not staying in it, including these unlikely permutations can slightly enhance the rate. We tested this hypothesis using the chevron (Figure \ref{fig:paretoCollection}(b)) as an example. If we restrict the target state to the 2-type permutation shown in the figure, the Pareto front reduces to a single point, 
supporting the hypothesis. However, the same is not true for the triangle (Figure \ref{fig:paretoCollection}(a)); we still find a near-vertical front when restricting the target to  the three particle type permutation shown in the figure. 
A second possibility is that adding auxiliary weak bonds can increase the rate of forming a target, even if that bond is not present in the target state. We could not find numerical evidence for this possibility (SI Appendix$^\dag$, Section \ref{sec:BDconfirm}). A third is that this observation is a numerical artifact, since we compute only an approximation to the true Pareto front.

\subsection{Lattice polymer}

\begin{figure}[h]
\centering  
\includegraphics[width=1\linewidth]{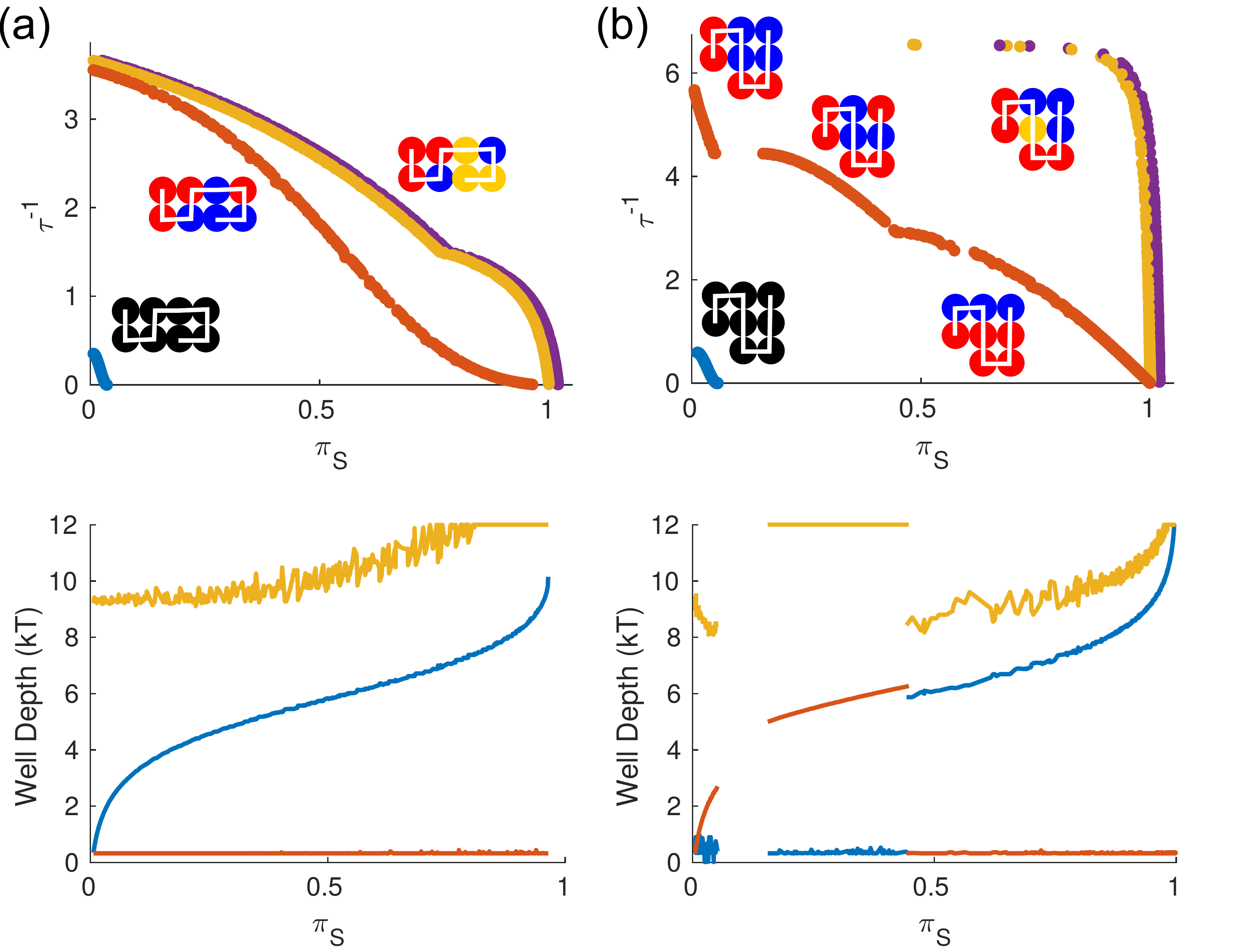}
\caption{Pareto fronts for two maximally bonded configurations of a lattice polymer. 
The information plotted is the same as in Figure \ref{fig:paretoCollection}. 
}
\label{fig:paretoLattice}
\end{figure}


As a final example we compute Pareto fronts for lattice polymers formed from a chain of 8 particles (SI Appendix$^\dag$, Section \ref{sec:latticeDetails}). 
Figure \ref{fig:paretoLattice} shows Pareto fronts for two different target states, each with a fixed permutation. For both target structures, the best folding is achieved at $m{=}3$, where the Pareto front is indistinguishable from the Pareto front for $m{=}8$. Many of the qualitative features of these fronts are the same the same as for disks, but a notable exception is the Pareto front for Figure \ref{fig:paretoLattice}(a) does not become nearly vertical, even for distinct particles (although the slope near $\pi_S{=}1$ approaches $\infty$.) 
We believe the absence of a nearly-vertical Pareto front is because our lattice model allows two bonds to form at once, so it captures some of the kinetic traps that get averaged out by our coarse-grained continuum model, an issue we discuss later.

\subsection{Pareto Front Sensitivity} \label{sec:sensitivity_main}

\begin{figure}[h]
\centering  
\includegraphics[width=1\linewidth]{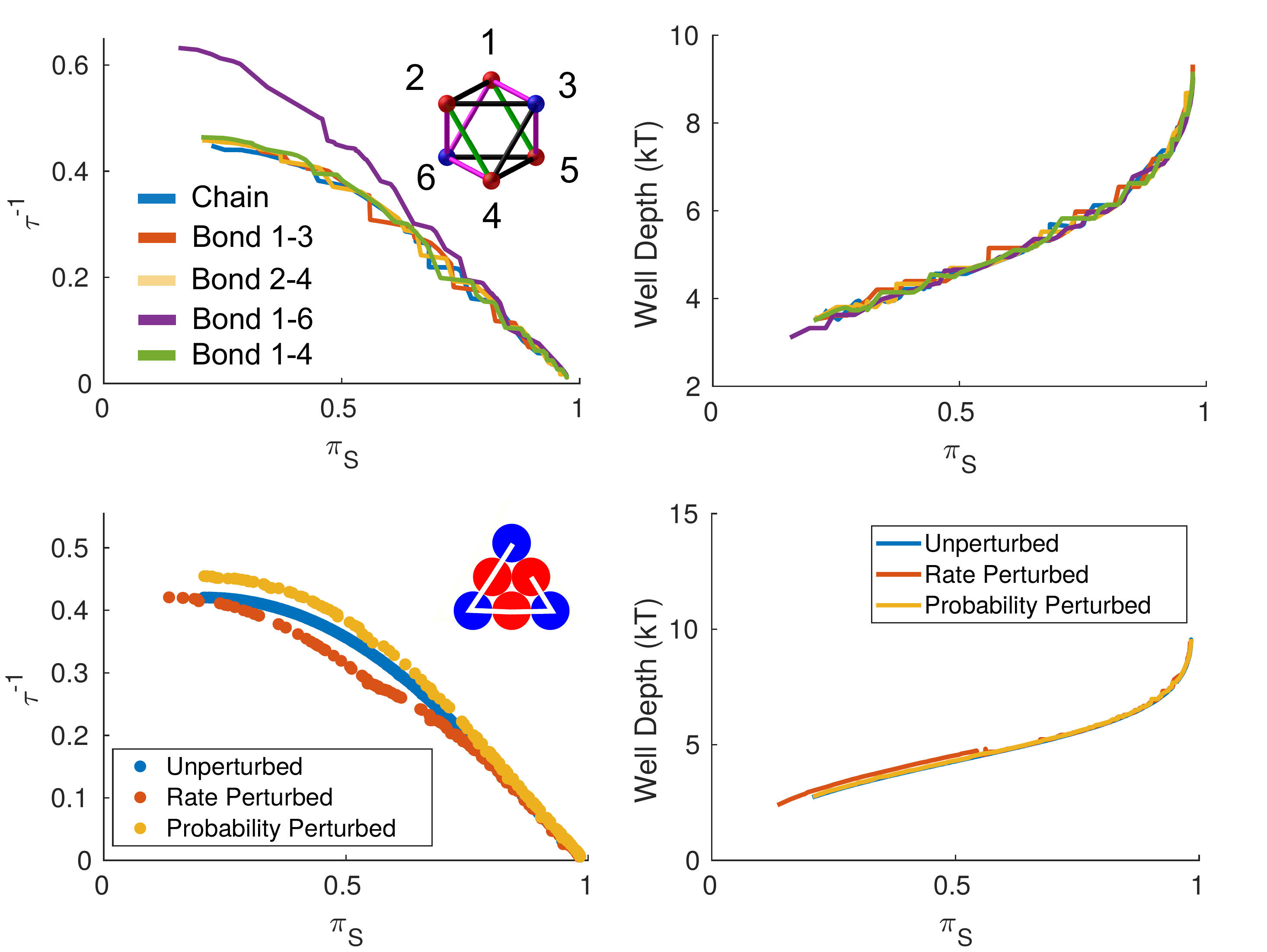}
\caption{Top Row: Pareto fronts and non-trivial bond energy parameterizations (AA interaction) for the octahedron state with two particle types, for various $1$-bond starting states. With the exception of the linear chain, all starting states are sampled from the equilibrium measure in that state. Starting with a mis-folded bond has nearly no effect, as the bond can quickly break, while starting in a loop increases the maximum rate, as this bond is the slowest in the assembly process. Bottom Row: Pareto fronts and parameterizations (AB interaction) for the two particle type triangle after perturbing model parameters. Exit rates out of coarse grained states are perturbed by adding Gaussian noise with standard deviation 20\% of each state's rate ($\sigma=0.2/\tau_{i}$.) Probabilities are perturbed by making each transition out of a state equally likely.  }
\label{fig:perturbMain}
\end{figure}

How sensitive are the Pareto fronts to the underlying assumptions of the model, such as the initial condition, estimation of free energies, model for transition rates, choice of pair potential and bond cutoff, neglect of hydrodynamic interactions, etc? We computed Pareto fronts for a small collection of examples under two types of perturbations: one which alters the transition rates, and another which considers different initial conditions (SI Appendix, Section \ref{sec:sensitivity}.) Figure \ref{fig:perturbMain} shows that while the numerical values of the rates along the Pareto fronts do change, their qualitative shape does not, and, remarkably, the bond parameterizations along the Pareto fronts appear insensitive to any of these perturbations. These findings suggest that the Pareto optimal interactions are mainly a function of the system's connectivity network, i.e. its topology, rather than its detailed kinetics.

Such a finding implies that strategies that attempt to minimize the number of accessible pathways to mis-folded states, such as was done for a minimal, $3$-level toy model of self-assembly \cite{whitelam_microscopic_reversibility}, may successfully avoid kinetic traps, without detailed knowledge of the folding kinetics. 




\section{Chiral traps: the problem with short-ranged isotropic interactions}\label{sec:chiral}

\begin{figure}[h]
\centering  
\includegraphics[width=0.6\linewidth]{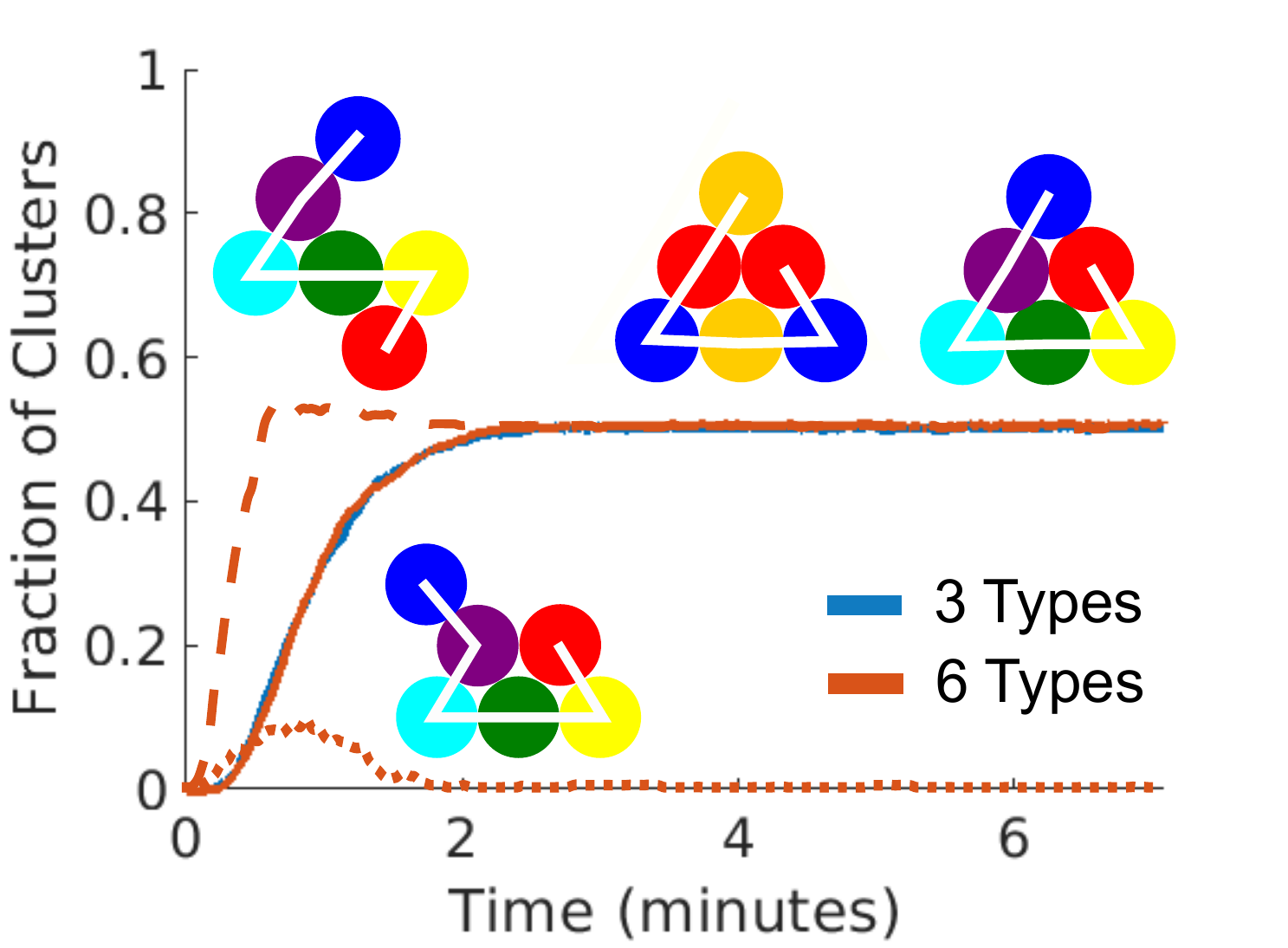}
\includegraphics[width=0.3\linewidth]{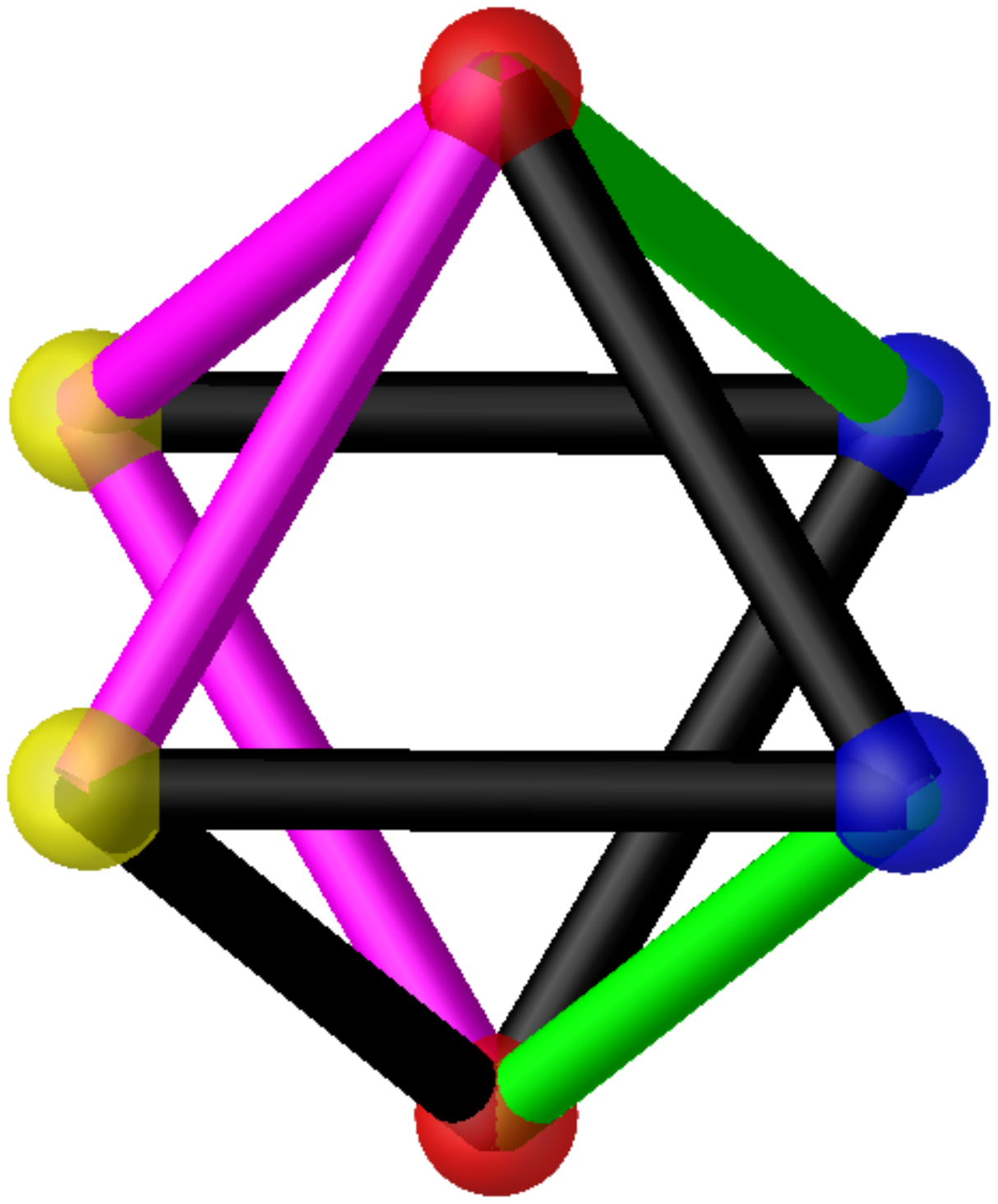}
\caption{Chiral traps prevent efficient assembly. 
Left: Brownian dynamics simulations of polymers of 6 disks, with 3 types (ABCBAC) and 6 types. 
The yield stagnates at about 0.5 in both cases, because half the trajectories get stuck in a kinetic trap (dashed line), which requires breaking a strong bond to form the triangle. 
The kinetic trap has the same adjacency matrix as an on-pathway cluster (dotted line), so it is impossible to remove this trap using short-ranged isotropic interactions. 
%
All yields were calculated from 400 trajectories initialized as a linear chain. 
The three-type chain had energies $E_{AA}{=} E_{AC} {=} 12 kT$, while the 6-type chain set all target energies to $12 kT$; all other interactions were $0.1 kT$. 
Right: The floppy kinetic trap for the octahedron, with the strong AB and AC bonds in green and pink respectively, and backbone bonds in black. This cluster has the same adjacency matrix as a cluster where the frontmost yellow and blue particles are interchanged, which can then form strong blue-yellow bonds with their neighbours. 
}
\label{fig:triangleBD}
\end{figure}

We encountered a problem when we tested whether the design parameters found by our genetic algorithm, lead to fast folding in Brownian dynamics simulations. 
Sometimes the Brownian dynamics simulations agreed with our predictions of fast folding, 
but sometimes the simulations got stuck in kinetic traps. For example, Figure \ref{fig:triangleBD} shows Brownian dynamics simulations of a triangle with 3 particle types and interactions found by the algorithm to have no tradeoff. 
The yield stagnates at about 0.5; it will eventually reach 1 but only over a much longer timescale. The other 50\% of the trajectories are stuck in a floppy state, which requires breaking a strong bond to leave. This floppy state has the same adjacency matrix as a floppy state which is on the folding pathway to the triangle. 
 Our coarse-grained model does not distinguish clusters with the same adjacency matrix: it lumps the kinetic trap and the on-pathway cluster into the same state, effectively allowing the system to tunnel between the two configurations, and thereby removing the energy barrier for leaving the trap. This problem occurs even for distinct particles: the kinetic trap always exists for the original, continuous system, even if it is removed in our coarse-grained model. 
 
 This problem also occurs for the octahedron, although more mildly; about 1-2\% of the trajectories got stuck in a floppy trap with the same adjacency matrix as an on-pathway cluster, shown in Figure \ref{fig:triangleBD}. This same floppy trap has been observed experimentally in DNA coated colloids assembling from a gas, with interaction energies similar to the ones we identified as optimal \cite{collins2014self}.
 
 We call such kinetic traps \emph{chiral traps}, because there is no way to distinguish them from on-pathway clusters given only the matrix of particle contacts. Chiral traps pose a fundamental problem for self-assembly or folding using short-ranged isotropic interactions. 
 There is no way to remove these traps using interactions that do not depend on particle orientation, even if particles are all distinct. Indeed, such traps have been observed in simulations of dozens of distinct particles \cite{zoranaSizeLimits}.
 
 Yet, all is not lost -- the success of our coarse-grained model in finding vertical Pareto fronts, suggests that were the chiral traps to be removed, or distinguished from their on-pathway counterparts, one could achieve fast, reliable assembly. One way to do this is with particles with patchy interactions \cite{zhang2004self,romano2010phase,Chen:2011be,patchy}. 
It could be that patchy particles have qualitatively similar Pareto fronts as with our coarse-grained model. This idea is supported by computational studies that observe high fidelity yields of icosahedra using patchy particles \cite{doye_patchyIcosahedra}; yield reductions were due primarily to aggregation effects, which are not present in our system, instead of mis-folding effects.
 %
 %
We also remark that the coarse-grained model is successful within the context of its own assumptions. Figure \ref{fig:triangleBD} shows that simulating a chain of distinct particles, gives the same yield curve as a chain with 3 particle types. Therefore, the model can identify when assembly is as efficient with $m$ particle types as it is with distinct particles.

 \section{Extension to Larger Systems} \label{sec:extension}
 
 So far we have computed Pareto fronts for systems whose state space can be fully enumerated, and we furthermore lumped states sharing the same adjacency matrices so that states can be easily identified \cite{Zappa:2019id}. 
 This allowed us to efficiently evaluate measures of thermodynamic stability and kinetic accessibility, but it will not be feasible for larger systems whose state spaces cannot be fully enumerated. 
 
 
A straightforward idea for handling non-enumerable systems would be to simulate them, and estimate the objectives from an ensemble of trajectories.  We do not expect this to be feasible -- first, because computing a Pareto front requires evaluating the objectives thousands of times, 
and second because many of the parameter values of interest lead to extremely slow rates of folding, which also imply very long simulation times. However, there is flexibility in choosing the objectives; 
other functions might be just as (or more) informative, yet possible to evaluate cheaply and accurately by simulation. 

\begin{figure}[h!]
\centering  
\subfigure[]{\includegraphics[width=0.55\linewidth]{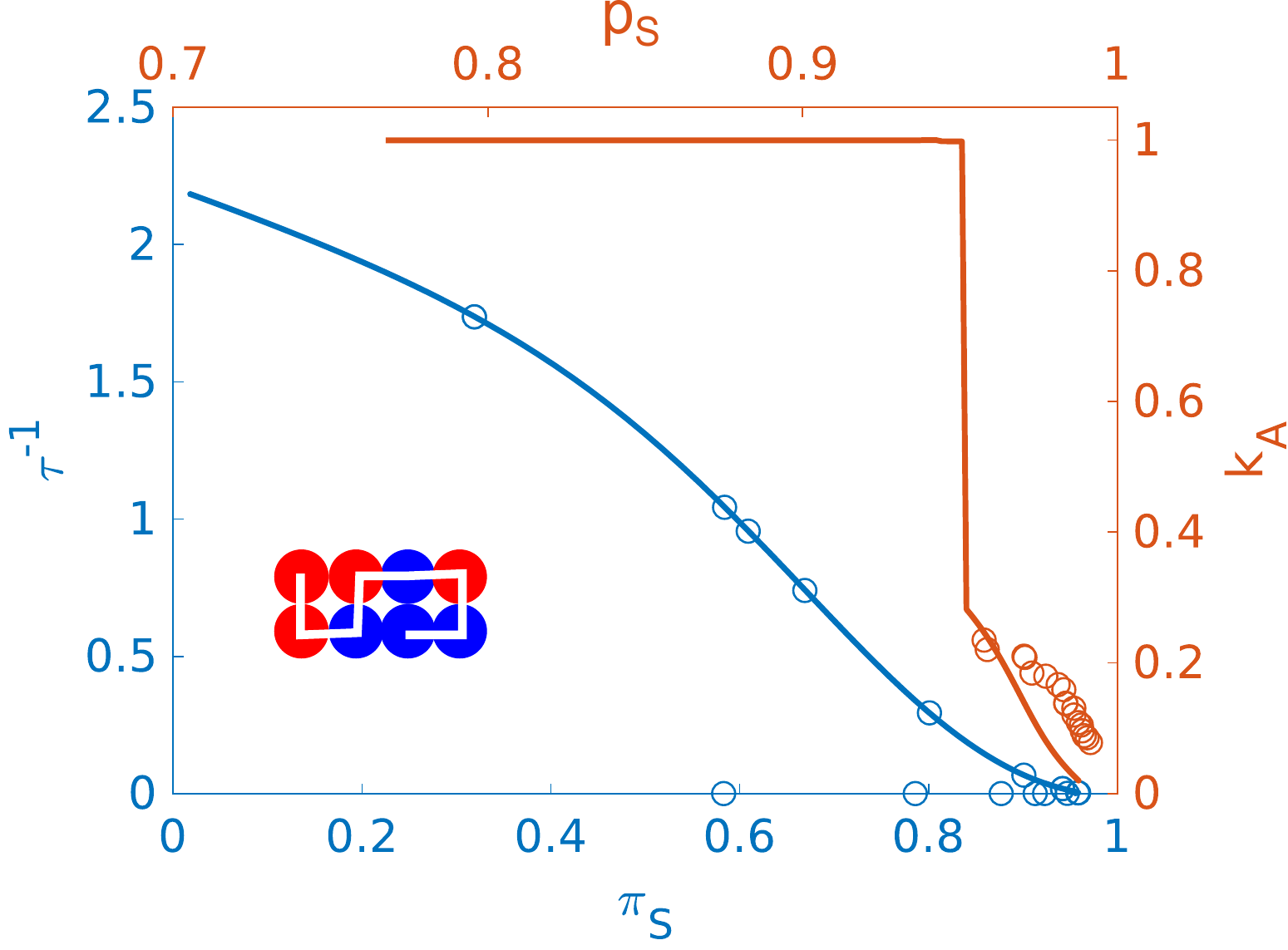}
\label{fig:samplingN8}
}
\subfigure[]{\includegraphics[width=0.33\linewidth]{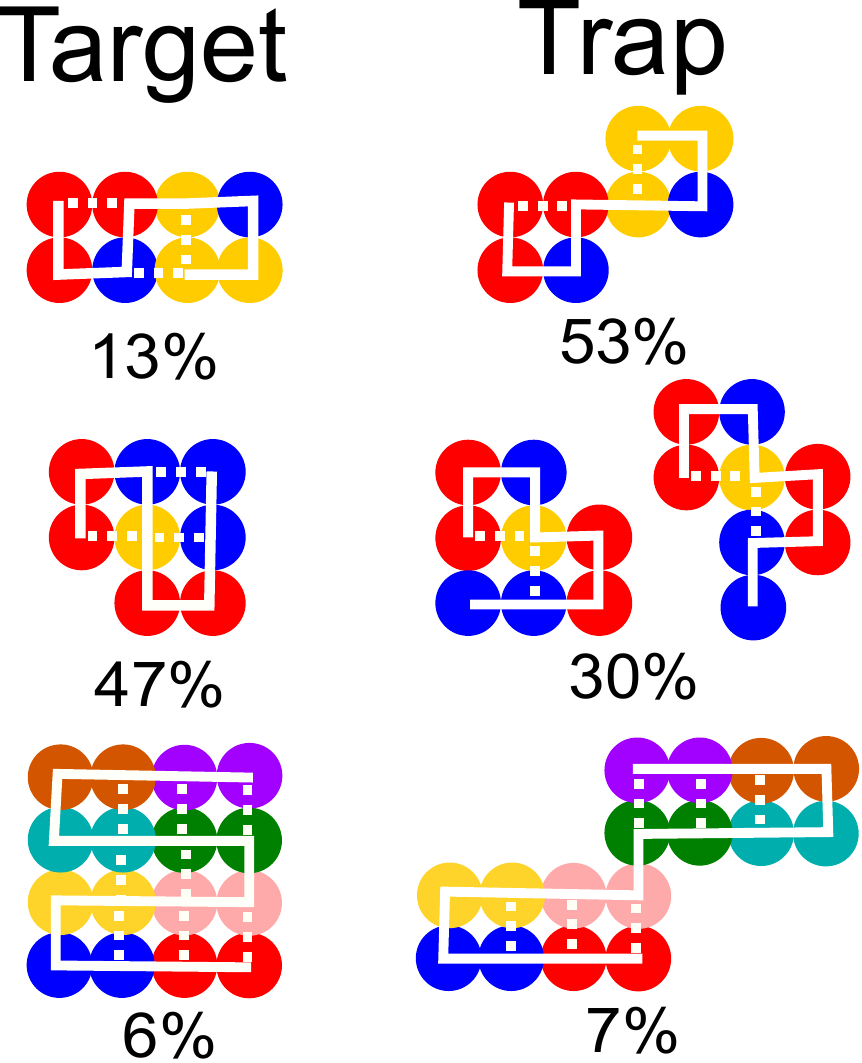}
\label{fig:latticeTraps}
}
\caption{
(a) Mapping of the Pareto front for the $N=8$ particle rectangle state on a lattice in the original objectives $(\pi_S, \tau^{-1})$ (solid blue curve) to the new objectives $(p_S, k_A)$ (solid red curve), computed with the coarse-grained model. The genetic algorithm was then applied to the new objectives by sampling, using $P=100$, $p=0.25$, $T=500$, $t_{\text{trap}}=300$ (see SI Appendix$^\dag$, Section \ref{sec:Larger} for details), $200$ generations, and $500$ samples to estimate averages. The non-dominated portion of the population is shown as unfilled red circles. The parameters for the red circles were extracted, and plugged into the coarse-grained model to evaluate the original objectives (blue circles), which can be seen to still lay mostly on the original Pareto front. 
(b) Yields of the target state and most common kinetic trap, estimating using $5000$ samples, for the two $N=8$ ground states at $T=5000$, as well as a square of $N=16$ particles at $T=50000$, using the Pareto optimal parameters from sampling. Solid white lines denote the backbone, and dashed white lines denote bonds with $E=12 kT$. Note that the contact matrix for the trapped states are consistent with the target state, showing the kinetic traps are chiral traps.
}
\end{figure}

We have tentatively explored a selection of other objectives. One idea that appears promising, is to replace $\pi_S$ with $p_s$, an average energy barrier for leaving the state, and $\tau^{-1}$ with $k_A$, a measure of the energy of mis-folded bonds when a trajectory first becomes trapped (see SI Appendix$^\dag$, Section \ref{sec:Larger} for details). These objectives may be estimated reasonably efficiently from an ensemble of trajectories. 
Figure \ref{fig:samplingN8} shows that for a lattice polymer of size $N=8$, the Pareto front computed for these new objectives, gives bond energies that lie mostly on the Pareto front for the original objectives. This shows the parameterization of the Pareto front is relatively unchanged by this change of objectives, a promising result. We then ran our genetic algorithm for a lattice polymer of size $N=16$, a system big enough that enumeration is impossible, and obtained converged Pareto optimal parameters (SI Appendix$^\dag$, Section \ref{sec:Larger}.) However, for both lattice polymers, the Pareto optimal parameters do not give a good yield of the target structure: Figure \ref{fig:latticeTraps} shows that the finite time target structure yield is significantly eroded by the presence of chiral traps. 
Since chiral traps contain no mis-folded bonds, they do not contribute to $k_A$, and thus are undetected by our sampling. By inspection, we see these trap states could also be eliminated by introducing directional bonds.

\section{Discussion}

We computed Pareto fronts describing the tradeoff between equilibrium probability and rate of folding, for small polymers of spheres and disks with programmable interaction energies. We showed that a genetic algorithm can compute these Pareto fronts, and can handle the discrete optimization problem of determining particle labels.  
This approach makes no assumptions on the intrinsic timescales of the assembly (in contrast to approaches which optimize yield at a fixed a final time), allowing the assembly timescales to emerge naturally through our optimization algorithm, and, it can identify collections of optimal parameters in a high-dimensional parameter space. 

A key approximation that let us efficiently evaluate the equilibrium probability and rate of folding, was to first construct a coarse-grained model which approximates the dynamics as a Markov chain on the set of adjacency matrices. 
The main limitation of this approximation is that it removes some kinetic traps for isotropic particles that sometimes prevent a real system from assembling efficiently. We believe other possible limitations are negligible -- indeed, we found 
the design parameters on the Pareto front are remarkably insensitive even to large perturbations. 


We exhaustively calculated the Pareto fronts for the ground states of small polymers with every possible number of types of particles, for this coarse-grained dynamical model. Apart from being useful as a design tools, these Pareto fronts give qualitative insight into what determines the thermodynamic-kinetic tradeoff. 
A striking observation was that for nearly all target states, there was a critical number $m_*$ of particle types, less than the total number of particles, that led to a vertical or point-like Pareto front. When this happens there is no tradeoff; one may choose interactions such that the polymer folds rapidly to a low-energy target state.

Our philosophical approach to calculating Pareto fronts is not limited to polymers of spherical particles, but could be adapted to handle particles assembling from a gas, particles with orientation-dependent interactions, particles with different shapes, etc; additional objectives could also be considered, resulting in Pareto fronts that are hypersurfaces with dimension one less than the number of objectives. 
Our technical approach to coarse-graining the configuration space, however, is limited to systems small enough that one can enumerate all the possible ways to put particles in contact. The main challenge in extending our results will be to efficiently evaluate measures of thermodynamic stability and kinetic accessibility, in systems where this enumeration is no longer feasible. 
We tinkered with other objectives, which seemed promising when evaluated for small lattice polymers, however we were discouraged by the increasing number of chiral traps that occur for larger lattice polymers, which prevent the system from ever folding efficiently. 
Any further work in this direction must consider non-isotropic interactions. 



We were motivated by the puzzling success of protein folding, and suggest that a steep Pareto front is a signature of a ``funnelled'' energy landscape. This funnel metaphor has proven invaluable in studying how proteins navigate their complex energy landscape to efficiently fold into their native states, but it has not yet found use as a practical design tool in materials science. Framing the design problem in the language of multi-objective optimization, as we have done, and building upon existing tools to address the tradeoffs between objectives, may provide an alternative method to design complex self-assembling systems given realistic experimental constraints. 
A fascinating question would be to 
understand how the critical number $m_*$ of particle types grows with system size or target structure complexity. 
Information theory \cite{Soloveichik:2007ey,demaine2011one} combined with statistical mechanics considerations may give additional insight into this question. 
Will this number always be less than 20, as for proteins?

\section{Conflicts of Interest}
There are no conflicts of interest to declare.

\section{Acknowledgements}
The authors are grateful to Eric Vanden-Eijnden for introducing them to this problem. 
This work was supported by the Research Training Group in Modeling and Simulation funded by the National Science 
Foundation via grant RTG/DMS -- 1646339, and by the United States Department of Energy under Award no. DE-SC0012296. 
M. H.-C. acknowledges support from the Alfred P. Sloan Foundation. 

\section{Data Availability Statement}
The raw data and scripts to make figures, as well as all the code to generate our data is available on Github \cite{cpfold_github}.

\bibliography{pub_refs} 

\section{Appendix}

\subsection{Equilibrium Probability Calculations}\label{sec:EqProb}

In equilibrium, the probability of observing a system in a given micro-state, $x$, is denoted $p(x)$, and is given by the Boltzmann distribution,
\begin{equation} \label{boltzmann}
    p(x) = \frac{1}{Z} e^{-\beta U(x)}.
\end{equation}
Here $Z$ is a normalizing constant called the \emph{partition function}, $\beta$ is the inverse temperature, and $U(x)$ is the potential energy of the micro-state $x$. We consider a chain of $N$ particles of unit diameter $d=1$, in two or three dimensions, with short-ranged, isotropic, pairwise interactions, and we assume the backbone interactions are so strong they do not break. When the interaction range is short enough, a good approximation is to treat this system in the \emph{sticky limit}, in which the range of the interactions goes to zero, i.e. particles interact only when they are in direct contact. Each micro-state $x$ can be assigned and grouped into a macro-state by its adjacency matrix. We will refer to these macro-states as \emph{clusters}. 

The probability of observing the system in cluster $j$ in equilibrium, $\pi_j$, can then be computed by integrating (\ref{boltzmann}) over the appropriate space. We let $C_j$ denote the set of all micro-states consistent with cluster $j$; including all rotations, translations, reflections, and deformations along internal degrees of freedom. We find
\begin{equation} \label{eqJ}
    \pi_j = \frac{1}{Z} \int_{C_j} e^{-\beta U(x)}dx = \frac{Z_j}{Z},
\end{equation}
where $Z_j$ is the contribution to the total partition function from cluster $j$. Evaluating the integral (\ref{eqJ}) in the sticky limit is quite involved; the calculations can be found in \cite{HolmesCerfon:2013jw,freeEnergySticky}, which can be adapted to consider assembly from a gas or from a polymer. The end result is
\begin{equation}\label{finalEq}
    \pi_j = \frac{z_j}{Z}\kappa^{b_j}, \quad Z = \sum_k z_k \kappa^{b_k}.
\end{equation}
Here $z_j$ is called the \emph{geometric} partition function for cluster $j$, and depends only on the geometric properties of the cluster such as the moment of inertia and symmetry number. All the dependence on the interactions is encapsulated by the \emph{sticky parameter}, $\kappa$, and $b_j$ denotes the number of bonds in cluster $j$. The sticky parameter is given by the partition function for a single bond,
\begin{equation} \label{stickyParameter}
    \kappa = \int_0^{r_c} e^{-\beta U(r)}dr,
\end{equation}
where $r_c$ is some cutoff distance beyond which the potential goes to zero. Using Laplace asymptotics to evaluate the integral (\ref{stickyParameter}), we find 
\begin{equation}\label{kappaest}
\kappa = \sqrt{2\pi} e^{-\beta U(d)} / \sqrt{\beta U''(d)},
\end{equation}
where $d=1$ is the minimum of the potential energy. When we estimate probabilities for an interaction potential that has finite range, such as the Morse potential considered for all the examples in the text, we use \eqref{kappaest} to convert the parameters for the potential into a sticky parameter. 

According to expression \eqref{finalEq} for $\pi_j$, the equilibrium probability only depends on $\kappa$, since $z_j$ is a constant for a given cluster. Therefore, as long as we know $z_j$ for all $j$, we can determine $\pi_j$ for any $\kappa$. Alternatively, if we can determine the value of all the $\pi_j$ for a given $\kappa_0$, we can determine equilibrium probabilities for any other $\kappa_1$, by performing a simple re-weighting and re-normalization. 
This reweighting works provided $\kappa_0,\kappa_1$ are close enough that there is sufficient overlap in their associated distributions; it was shown in \cite{HolmesCerfon:2020dk} that the reweighting works over fairly large range of sticky parameters for small clusters. 

The reweighting works as follows. Let $K_0^j = \kappa_0^{b_j}$ and $K_1^j = \kappa_1^{b_j}$. The true equilibrium distributions for these sticky parameter choices are
\begin{equation} \label{trueEqDist}
\pi_j^0 = \frac{z_j K_0^j}{\sum_n z_n K_0^n}, \quad \pi_j^1 = \frac{z_j K_1^j}{\sum_n z_n K_1^n}.
\end{equation}
Next, we construct an expression for $\pi_j^1$ in terms of $\pi_j^0$. We let
\begin{equation} \label{reweightExpression}
    \Pi_j^1 = \frac{\pi_j^0\frac{K_1^j}{K_0^j}}{\sum_n \pi_n^0\frac{K_1^n}{K_0^n}},
\end{equation}
and we show that $\Pi_j^1 = \pi_j^1$ by showing that their ratio is $1$. The calculation shows
\begin{align*}
    \frac{\Pi_j^1}{\pi_j^1} & = \frac{\pi_j^0 K_1^j}{K_0^j \sum_n \pi_n^0\frac{K_1^n}{K_0^n}} \frac{\sum_n z_n K_1^n}{z_j K_1^j}\\
    & = \frac{z_j K_0^j}{\sum_n z_n K_0^n} \frac{K_1^j}{z_j K_1^j} \frac{\sum_n z_n K_1^n}{1} \frac{1}{K_0^j \sum_n K_0^n \frac{z_n}{\sum_m z_m K_0^m }\frac{K_1^n}{K_0^n}}\\
    &= \frac{K_0^j}{\sum_n z_n K_0^n} \frac{\sum_n z_n K_1^n}{1}\frac{\sum_m z_m K_0^m}{K_0^j\sum_n z_n K_1^n}\\
    &= 1. 
\end{align*}
This shows that we can evaluate the equilibrium probabilities at any value $\kappa_1$ by knowing the equilibrium probabilities at some reference value, $\kappa_0$, and applying Equation (\ref{reweightExpression}). All that remains is to compute the reference set of equilibrium probabilities. 

While it is possible to evaluate reference probabilities $\pi_j^0$ semi-analytically for the rigid ground states \cite{mengExp}, this is not possible for floppy states, so instead we estimate them by Monte Carlo sampling. We utilized the stratification sampler developed by Holmes-Cerfon \cite{HolmesCerfon:2020dk} to sample the system exactly in the sticky limit. This sampler proposes moves on a constraint manifold in which bonded particles are exactly unit distance apart, and can also propose to jump to a new constraint manifold, which corresponds to forming or breaking a bond. We only allow non-backbone bonds to break since we assume the backbone bonds are so strong they are unbreakable. We set $\kappa_0=2$ to quickly explore the entire state space. During a single run, we generate $5\times 10^6$ points, saving every fifth data point.  We then split the data sequentially into ten bins and construct  estimators  of  the  equilibrium  probability  of  each  state  in  each  bin. We compute error bars as the variance across the bins.  If we perform this procedure a few times, we can then combine these estimates, weighting them in such a way as to minimize the variance.  We stop when the largest relative error bound does not exceed $3\%$. This required repeating 4 times for 6 disks and 6 spheres, and 6 times for 7 disks. 


\subsection{Brownian Dynamics Simulations}\label{sec:BD}
We perform Brownian dynamics simulations of a chain of colloidal particles to study the kinetics of folding. 
Let $\vec{X}_t$ be a vector denoting the coordinates of the center of $N$ particles of diameter $d$ as a function of time. We choose $\vec{X}_0$ such that the particles are aligned on the $x$-axis, spaced a distance $d$ apart. The positions are then evolved via the overdamped Langevin equation,
\begin{equation} \label{langevin}
    \frac{d\vec{X}_t}{dt} = -\frac{D}{kT} \nabla U(\vec{X}_t) + \sqrt{2D}\eta(t),
\end{equation}
where $D$ is the diffusion coefficient for a particle, $k$ is Boltzmann's constant, $T$ is the temperature, $U$ is a potential energy function, and $\eta(t)$ is a vector of independent white noises. The potential energy is computed as a sum over pair potentials between each particle. To model the short-ranged interactions of the colloids, we use a Morse potential with the form
\begin{equation}
    U_M(r) = E\left(e^{-2\rho(r-d)}-2e^{-\rho(r-d)}\right).
\end{equation}
Here, $r$ is the inter-particle separation, $E$ is the bond energy, and $\rho$ is a parameter governing the width of the potential. We use a constant $\rho=40$ in our simulations, which corresponds to an interaction distance of about $6\%$ of the particle diameter \cite{SHScont}. The bond energy is left as a design parameter. 

To keep the chain intact, we introduce a stiff harmonic potential between the initially adjacent particles, which replaces the previously described Morse potential. The potential has the form
\begin{equation}
    U_H(r) = \frac{k}{2}(r-1)^2,
\end{equation}
with spring constant $k=2\rho^2E_0$, to match the curvature of a Morse potential at the minimum. We set $E_0=14 kT$ for six particle systems and $E_0=12 kT$ for seven particle systems. With these choices, we observe that the backbone almost always stays intact, and discard trajectories in the rare cases in which the backbone breaks. 

This equation can be non-dimensionalized by scaling positions by the particle diameter, $d$, scaling time by an unknown parameter, $c$, and scaling the potential energy by $kT$. Re-using the same notation, the non-dimensional equation looks like
\begin{equation} \label{ndLangevin}
    \frac{d\vec{X}_t}{dt} = -\epsilon \nabla U(X_t) + \sqrt{2\epsilon}\eta(t),
\end{equation}
where $\epsilon = \frac{Dc}{d^2}$ is a dimensionless parameter. We perform our simulations at $\epsilon = 1$. 

We numerically solve Equation \ref{ndLangevin} using the Euler-Maruyama scheme \cite{gardiner2004handbook}. Due to the stiff potential, we use a time-step of $\Delta t = 5\times 10^{-6}$ to ensure stability of the numerical scheme. 

To convert our non-dimensional simulation time, $t_{\text{sim}}$, to physical time in a lab, $t_{\text{lab}}$, we determine the value of our scaling parameter, $c$, using experimental data. Experiments with  clusters of $d = 1.3\mu$m  colloids above a wall, showed that the diffusion coefficient  is approximately $D = 0.1\mu\text{m}^2/\text{s}$ for an isolated particle, and slightly smaller on average for diffusion along an internal degree of freedom, $D = 0.065 \mu\text{m}^2/\text{s}$ \cite{experimentColloidDiffusion}. Using these values, a lower bound for our time scaling is $c\approx 17$ seconds. The estimated lab time is then given by $t_{\text{lab}} = c t_{\text{sim}}$. 

Although this scaling is for colloids diffusing above a wall, we apply it to both 3-dimensional and 2-dimensional clusters of colloids indiscriminately. We do not expect the scaling to give quantitative agreement with any particular experiment, which would anyways require considering different diffusion coefficients for different internal degrees of freedom \cite{folding}, but we do expect it to give an estimate of the order of magnitude of the timescales involved. Regardless, the ratio of folding timescales we observe in simulations with different bond energies does not depend on the scaling we choose.  

To determine what state the system is in at time $t$, we construct an adjacency matrix, $A_t$, using the coordinates $\vec{X}_t$. Let $r^{ij}_t = \|\vec{X}^i_t-\vec{X}^j_t\|$ be the distance between particles $i$ and $j$ at time $t$. The adjacency matrix is then constructed such that $a^{ij}_t = 1$ if $r^{ij}_t < r_{\text{cut}}$ and $a^{ij}_t = 0$ otherwise. We used a cutoff value $r_{\text{cut}} = 1.04$. This adjacency matrix is then compared against a database of adjacency matrices, that was precomputed when we performed the equilibrium probability calculations as in Section \ref{sec:EqProb}.

When reporting cluster yields from our Brownian dynamics simulations, we smooth the yield curves by averaging over a moving window with size equal to $0.5\%$ of the total number of time points.

\subsection{Coarse Grained Dynamical Model}\label{sec:coarse}

\subsubsection{Overview}
 
Our model approximates the dynamics on the energy landscape as a Continuous Time Markov Chain (CTMC), where each node or state of the Markov chain represents the configurations of a cluster with a particular set of bonds, described by its adjacency matrix. We treat particles as distinguishable in this step, so we do not lump together clusters whose adjacency matrices are equivalent under a permutation of particle labels. We are interested in both the equilibrium probabilities and the dynamics of this CTMC. 
The equilibrium probability $\pi_i$ of state $i$ is given by integrating the Boltzmann distribution over the continuous set of configurations that have bonds associated with state $i$. We estimate this integral for a given set of bond energies, and then reweight the estimate to obtain equilibrium probabilities for other bond energies, as described in Section \ref{sec:EqProb}.  

The dynamics of the CTMC are fully specified by the rates $Q_{ij}$ of transitioning from state $i$ to state $j$. We approximate these using the maximium likelihood estimators we would obtain from a long trajectory of the continuous dynamics with infinitely strong bonds, combined with the principle of detailed balance. That is, we set $Q_{ij} = 0$ unless states $i,j$ are related by either adding or breaking a single bond. 
If a bond is added, i.e. state $j$ has the same bonds as $i$ plus one extra, then we choose $Q_{ij}  = P_{ij} / \tau_i$, where $\tau_i$ is the mean first-passage time to form a bond from state $i$, starting with the equilibrium distribution and conditioned on no other bonds breaking, and $P_{ij}$ is the probability that when this first bond forms, it is the one from state $j$. 
Notably, $Q_{ij}$ does not depend on the bond energies, it only depends on the diffusion coefficient and the shape of the manifold of configurations corresponding to state $i$. 
For the reverse transition, from state $j$ to a state $i$ with one bond broken, we set the rate using detailed balance: $Q_{ji} = Q_{ij}\pi_i / \pi_j$. This choice ensures that the equilibrium probability for the CTMC is still $\pi = (\pi_i)_i$. 
Details of how we estimate $P_{ij}, \tau_i$ are in the following section. 


We remark that it might be more appropriate to estimate the exit probabilities and rates $P_{ij}, \tau_i$ by starting from the quasi-stationary distribution, not the equilibrium distribution, a distribution that is appropriate for describing rare exits from metastable states; it would be  interesting to know whether this give significantly different estimates \cite{qsd}. 
We are also neglecting hydrodynamic interactions, which can affect dynamics quite significantly \cite{folding}. 

Given our rate matrix $Q=(Q_{ij})_{i,j}$ and its equilibrium distribution $\pi$, we may solve for many kinetic quantities using linear algebra. In this paper we consider the mean first passage time (mfpt) $\tau$ from the linear chain (state 0) to our target state. Let $S$ be the indices corresponding to all adjacency matrices that are identified as our target state. Define a vector of mean first passage times, $\vec\tau=(\tau_0,\tau_1,\ldots)$ such that
\begin{equation}
    \tau_i = E[\inf\{t\geq 0 \text{ such that } X_t\in A\}| X_0 = i].
\end{equation}
Each component $\tau_i$ gives the mfpt from state $i$ to a state in $S$. 
The mfpt from the linear chain (state 0) to $S$ is found by solving the system of linear equations \cite{norris_1997}
\[
(Q\vec\tau)_i = -1 \;\; (i\notin \{0\}\cup S), \quad
\tau_i = 0 \;\;(i\in S),
\]
for the vector $\vec\tau=(\tau_0,\tau_1,\ldots)$, and then setting $\tau=\vec\tau_0$. 

In our examples the rate matrices $Q$ are typically quite sparse and structured. This is because each state can only be connected to other states with one more or one fewer bonds, and these connections are further constrained by cluster geometry. We take advantage of this fact by using a sparse solver to invert these linear systems. For $N=6$ disks, we use a dense-LU factorization based solver. In this case, the system is small enough that using a sparse solver in unnecessary. For $N=7$ disks and $N=6$ spheres, we use a sparse-LU factorization based solver. All linear algebra calculations are performed using the Eigen C++ library \cite{eigenweb}.

\subsubsection{Parameter Estimation}
To construct the CTMC model, we need estimators for $\pi_i$, the equilibrium probability of state $i$, $\tau_i$, the mean first exit time out of state $i$, and $P_{ij}$, the probability distribution of which state $j$ forms when exiting state $i$, for all $i$. We  discussed how to estimate $\pi_i$ in Section \eqref{sec:EqProb}. We now address how to estimate $\tau_i$ and $P_{ij}$. 

The mean first exit time out of state $i$ is defined to be
\begin{equation}\label{mpft}
\tau_{i} = E\left[\min\{n\geq 0 \text{ such that } X_n \neq i\}|X_0 = i\right],
\end{equation}
the expected value of the first time a new state is reached, conditioned on starting in the equilibrium distribution of state $i$ and having no existing bonds break. One approach to estimating $\tau_i$ would be to sample $M$ trajectories, equilibrated in state $i$ and then evolved in time chunks $\delta t$, until the next state, $j$, is formed. Using this method would require significant computation in order to equilibrate each sample independently. Instead, we propose an estimator that only requires a single long trajectory, using all of the data along the way. 

Assume an equilibrated trajectory forms a bond at time step $n$. Then at step $1$ of the trajectory, we know it will take $n$ steps to exit state $i$. At step $2$, it takes $n-1$ steps. Continue until at step $n$ we know it takes $0$ steps to exit. Each of these samples is drawn from the equilibrium distribution and can be combined into an estimator for the mean first passage time. We get
\begin{equation}\label{mfptEst}
\widehat{\tau_i} = \frac{\delta t}{n}\left(1+2+\dots+n\right) =\frac{n(n+1)}{2n}\delta t
\end{equation} 
as an estimator after a single transition. We then apply a reflecting boundary condition (i.e. set the state back to what it was at time step $n-1$), reset the timer, and repeat the process, evolving the trajectory independently of the first $n$ steps. To evaluate the total contribution to the estimator after several exits, one only needs to record the exit times, $T_k$, for $k=1,\dots, M$, and the number of times each state $j$ is reached, $c_j$. After $M$ exits, the estimators can be written as
\begin{align}\label{mfptEstTotal}
\widehat{\tau_i} &= \frac{\delta t}{\sum_{k=1}^M T_k} \sum_{k=1}^M \frac{T_k(T_k+1)}{2}, \\
\widehat{P_{ij}} &= \frac{c_j}{M}, \\
\widehat{Q_{ij}} &= \frac{\widehat{P_{ij}}}{\widehat{\tau_i}}.
\end{align}
Note that $\sum_{k=1}^M T_k$ is simply the total number of time steps sampled. Since the rate of bond formation depends only on the underlying diffusion and not the particle interactions, these estimates only need to be performed once, to sufficient accuracy. 

For each state $i$, we generated six trajectories in parallel using Brownian dynamics simulations, checking for bond formation at time chunks $\delta t = 0.01$. Each trajectory consisted of $16667$ exit samples, for a total of about $100,000$ samples for each state. We use the six trajectories to compute error bars on each $\tau_i$ estimate. We repeat the sampling procedure until the relative error bars are less than $8\%$ for each state $i$, which took us three to four applications of the algorithm. 

This estimation can be excessively time consuming when using Brownian dynamics simulations, if there are a lot of clusters. We use Brownian dynamics simulations to generate trajectories in the case of $N=6$ disks. For the larger systems, we further approximate the dynamics with the stratification sampler described in Section \ref{sec:EqProb}. Although this is a Monte Carlo sampler, it proposes small, local moves, which can move along any of a cluster's internal degrees of freedom, and therefore we expect it approximates overdamped dynamics for a small enough timestep (though this has yet to be shown rigorously.) 
This sampler can be modified to account for reflecting boundary conditions by removing proposal moves that jump to new constraint manifolds. Doing this will sample the equilibrium distribution of state $i$, and we denote a bond formation event when two particles are within a cutoff distance. The result is a sequence of configurations, but with no temporal information. To get approximate dynamical information from this scheme, we make use of the mean-square displacement property of a Brownian particle. For a one dimensional Brownian motion, the mean-square displacement is related to time via the formula $\langle (X(t)-X(0))^2\rangle = 2Dt$, where $D$ is the diffusion coefficient. For our proposal moves, we use an isotropic Gaussian over the tangent space of the current point on the manifold. If we choose the standard deviation, $\sigma$, of this proposal to be small, most moves will be accepted, and the mean-square displacement will be proportional to $\sigma^2$. We use $\sigma = 0.15$ in our simulations. Based on this, we introduce an approximate artificial time for each MC step as $\Delta t_{MC} = \sigma^2/2$. Any constant multiplicative factors that may be missing will multiply each estimated rate in the same way, meaning this method should approximate the rate matrix up to a multiplicative constant. 
We again generate six trajectories in parallel, each consisting of $16667$ exit samples, and compute error bars across the trajectories.

%
%

\subsection{Genetic Algorithm Details} \label{sec:GAdetails}

\begin{figure}[h!]
\centering  
\subfigure[]{\includegraphics[width=0.45\linewidth]{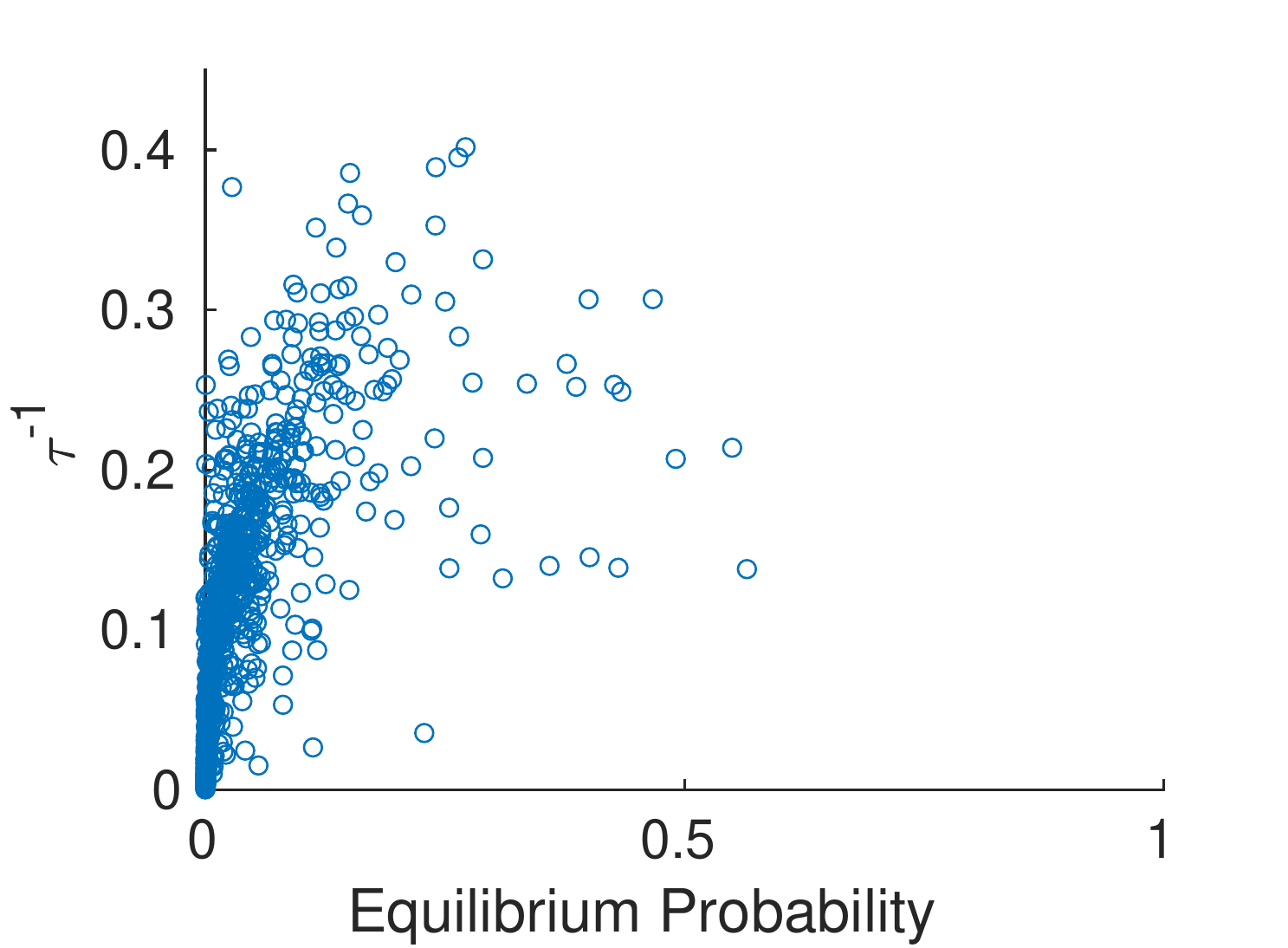}
}
\subfigure[]{\includegraphics[width=0.45\linewidth]{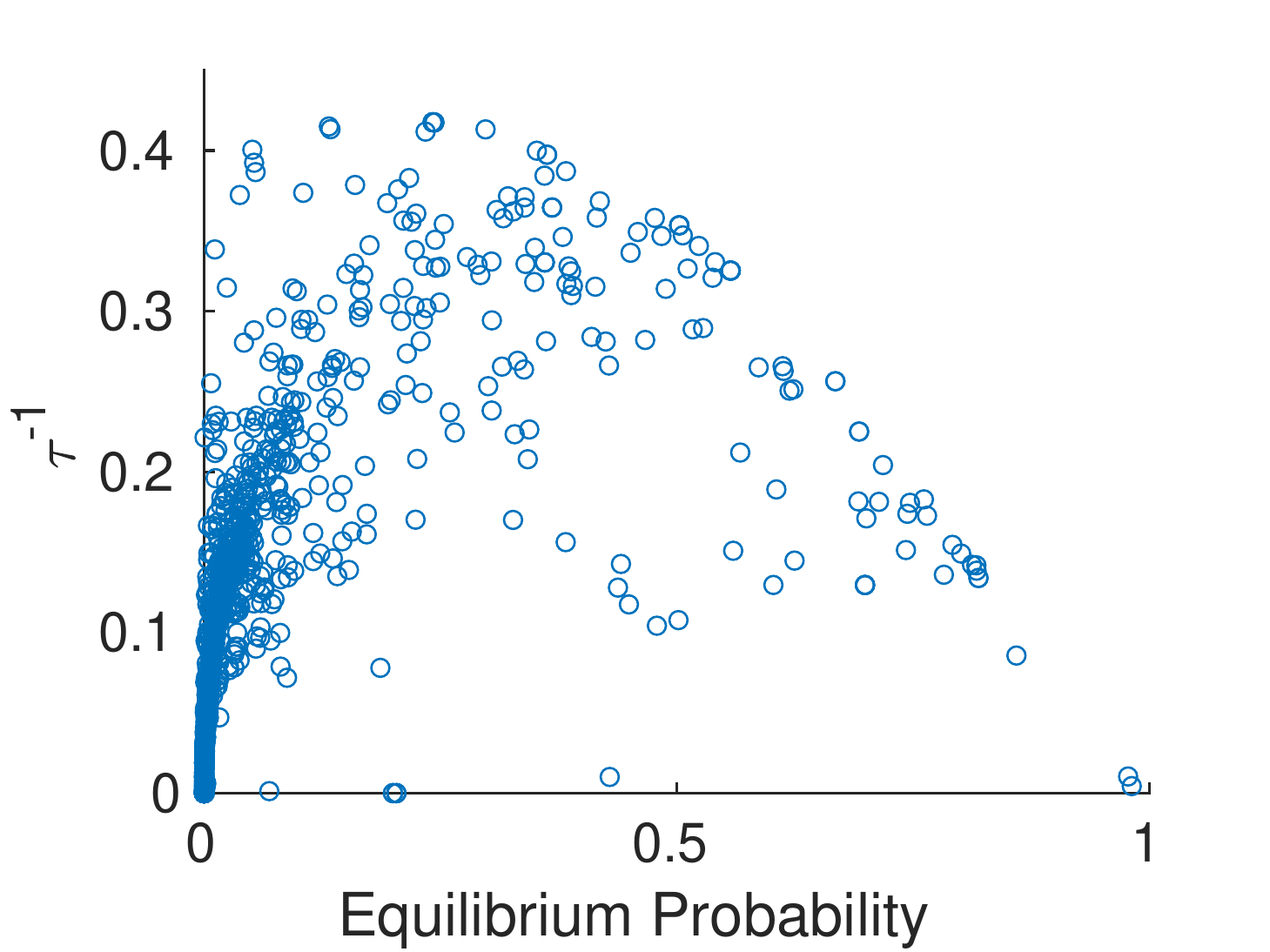}
}
\subfigure[]{\includegraphics[width=0.45\linewidth]{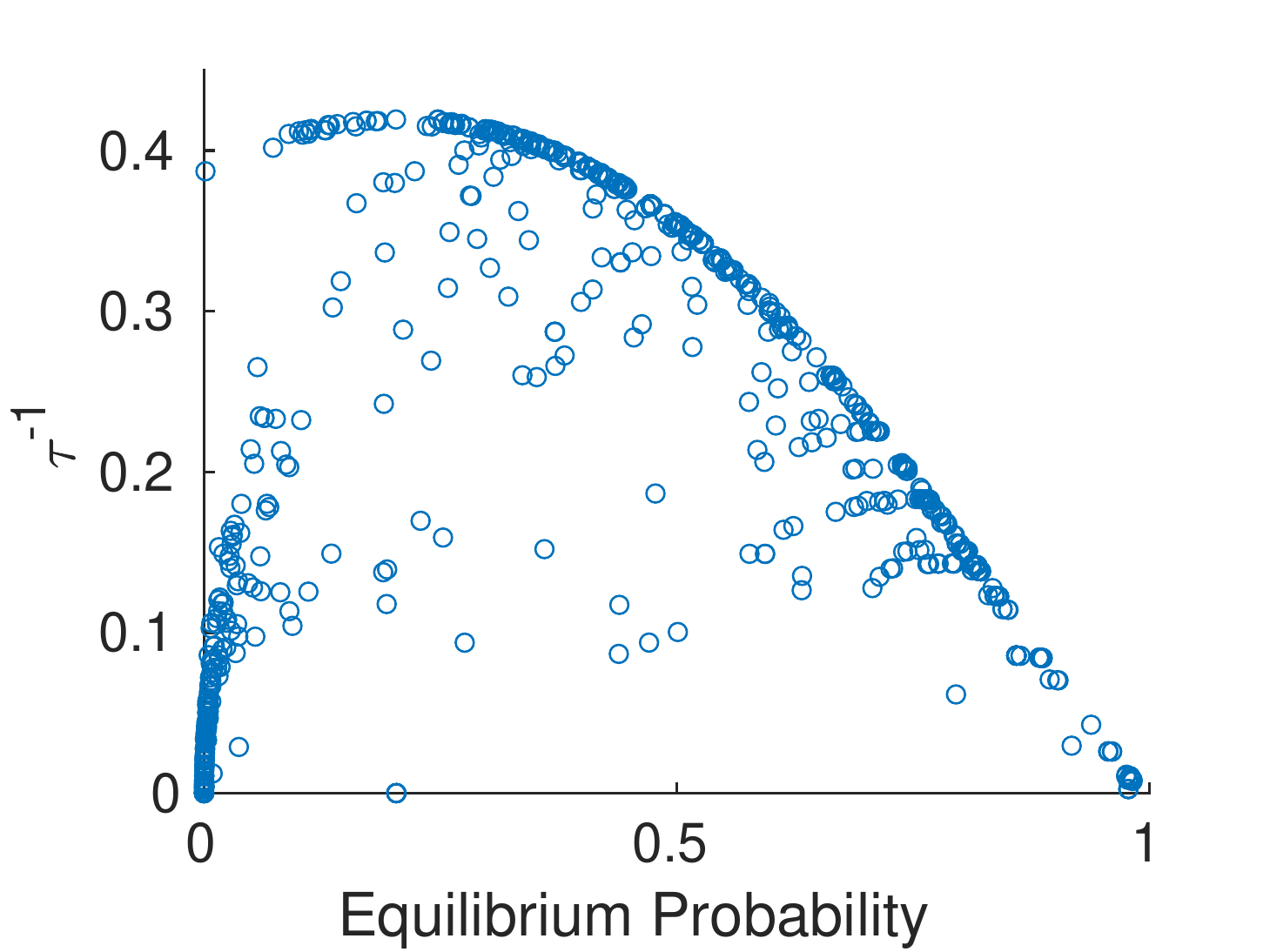}
}
\subfigure[]{\includegraphics[width=0.45\linewidth]{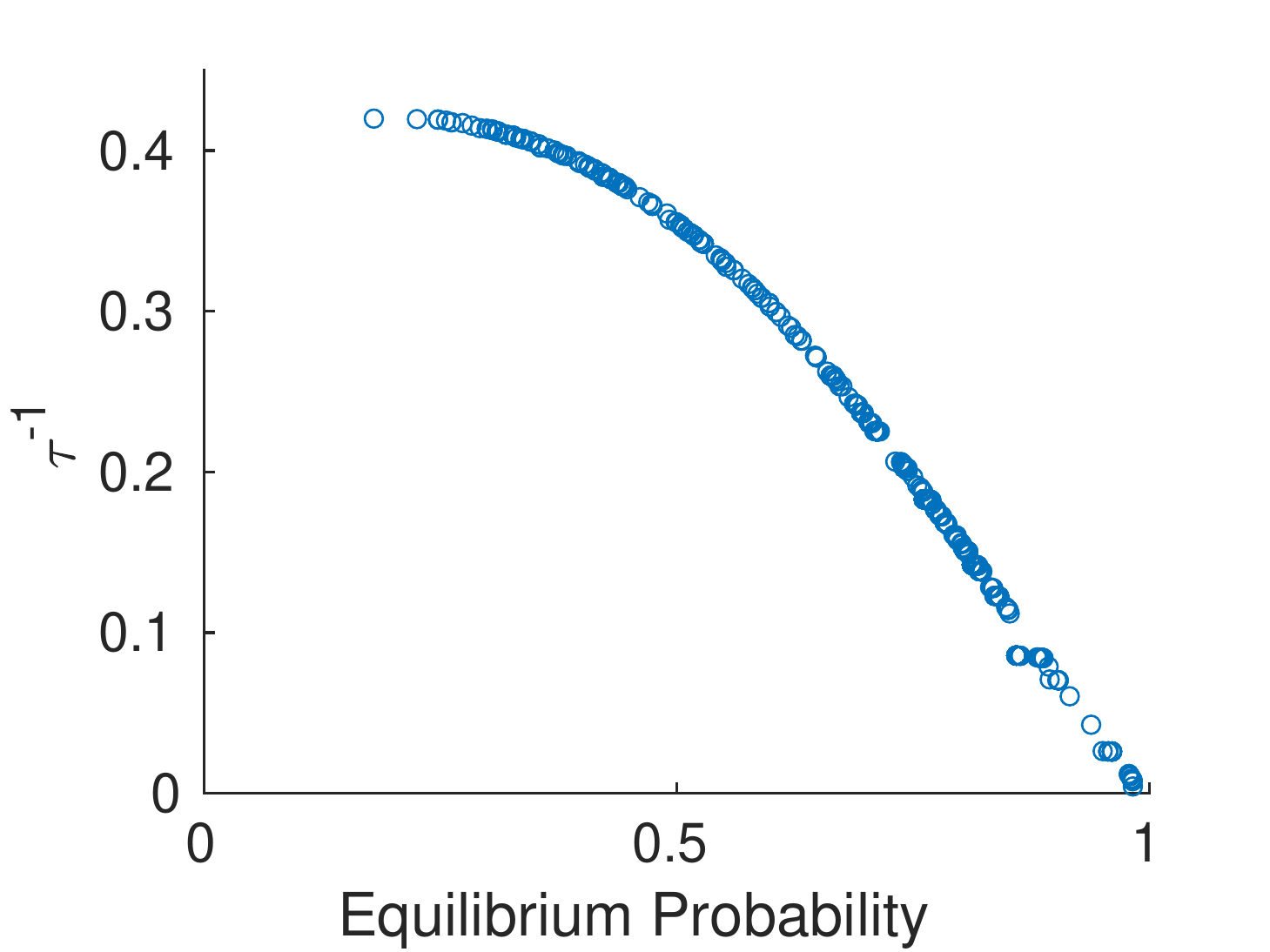}
}
\caption{Time evolution of the population from an application of the genetic algorithm in identifying the Pareto front for the self assembly of a triangle from an ABABAB chain. The generations shown are $1$, $7$, $15$, and $25$, in (a)-(d), respectively. This run used $P=1000$ members, $p=50\%$ as the cutoff for the mating pool, $r=10\%$ mutation chance, and $p_M = 40\%$ maximization chance with a threshold $E^* = 8$. }
\label{fig:genetics}
\end{figure}

The basics of our genetic algorithm were outlined in the main text. In addition to the cross-over operation, and re-sampling from mutation, we also implement a `maximizing' mutation. The idea behind this is that the optimal design parameters typically consist of at least one interaction being as strong as possible. Thus, to improve the quality of solutions and convergence speed, we build this into a mutation. If a given energy, $E$ is greater than some threshold $E^*$, then with probability $p_M$, we set the energy to the maximum value, $E_M$. An example showing the convergence of the population to the Pareto front is shown in Figure \ref{fig:genetics}. 

Next, we discuss how we compute non-dominated points, and our dominated points metric. There are divide-and-conquer approaches, but we simply perform an $O(P^2)$ brute force method. For each point, $i$, we compare its objective values to each other point, $j$, and count how many points $j$ have greater values in both objective functions, which we call $D_i$.  We do this for each point, $i$, and then sort the population from least to greatest $D_i$. If a point has $D_i = 0$, this point is Pareto optimal for the current generation, and is carried over to the next generation with probability $1$. The remainder of the population is filled by mating.

\subsection{Sensitivity Analysis}\label{sec:sensitivity}

\begin{figure}[h!]
\centering  
\includegraphics[width=0.99\linewidth]{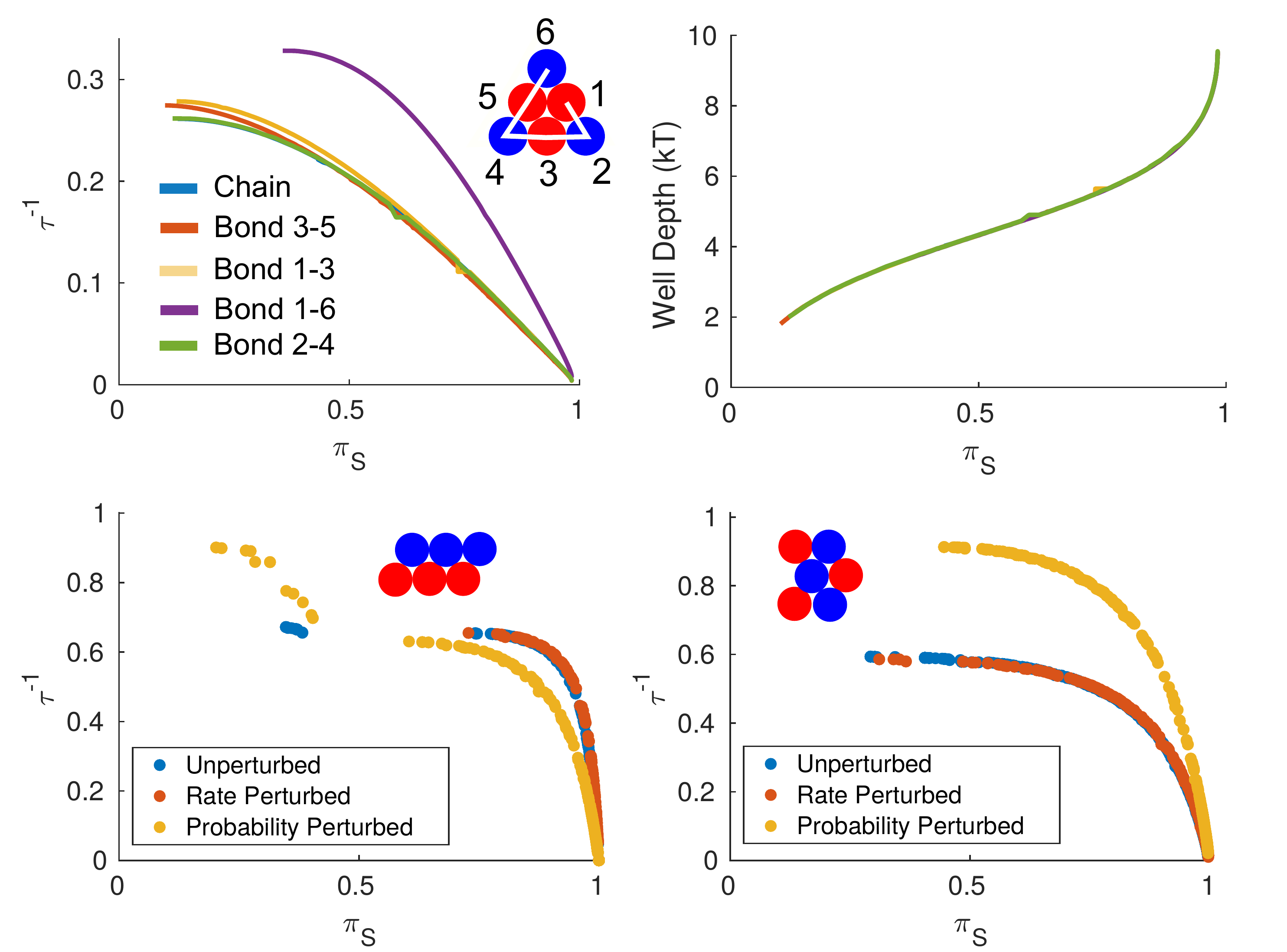}
\caption{Top Row: Pareto fronts and non-trivial bond energy parameterizations for the triangle state with two particle types, for various $1$-bond starting states. With the exception of the linear chain, all starting states are sampled from the equilibrium measure in that state.  Bottom Row: Pareto fronts for the other six-disk clusters after perturbing model parameters. Exit rates out of coarse grained states are perturbed by adding Gaussian noise with $\sigma=0.2/\tau_{i}$. Probabilities are perturbed by making each transition out of a state equally likely. 
}
\label{fig:perturbSI}
\end{figure}

In our analysis, we perturb the model kinetics in two ways to probe how sensitive the Pareto fronts are to the exact parameters of the model. One test is to add random noise to the exit rate of each state. To do this, for each state $i$, we set the perturbed exit rate to be $\hat{r}_i = \frac{1}{\tau_i} + \sigma N$, where $\tau_i$ is the mean first exit time out of state $i$, $N$ is a standard normal random number, and $\sigma$ is set to $0.2/\tau_i$, so typical perturbations in the rate are within about $20\%$ of the estimated value. Our second procedure involves perturbing the transition probabilities. Instead of using the measured values, we impose a uniform distribution over the set of possible transitions, which is a drastic change. 

In addition to perturbing model parameters, we also gauge the Pareto front's sensitivity to the assumed initial state. Instead of using the linear chain as the initial state, we repeat our calculations assuming the system starts in various $1$-bonded configurations, drawn from the equilibrium distribution in that connectivity state.  We chose some example configurations that were consistent with the target state, and others with mis-folded bonds. 

Figure \ref{fig:perturbSI} shows the results of these perturbation on the Pareto fronts for the system of six disks using an ABABAB chain. We find that perturbing the rates in the way we do has a marginal effect on the Pareto fronts. Perturbing the transition probabilities can have a large effect on $\tau^{-1}$ but the Pareto fronts still have the same qualitative behavior. The same is true of the choice of initial state; most of the Pareto fronts are very similar with the exception of starting in a loop, which assembles at a higher rate but with a qualitatively similar Pareto front.  The bond energy parameterizations along the Pareto front are remarkably insensitive to all of the perturbations we tested. 



\subsection{Confirming Observations With Brownian Dynamics Simulations}\label{sec:BDconfirm}

\begin{figure}[h!]
\centering  
\subfigure[]{\includegraphics[width=0.45\linewidth]{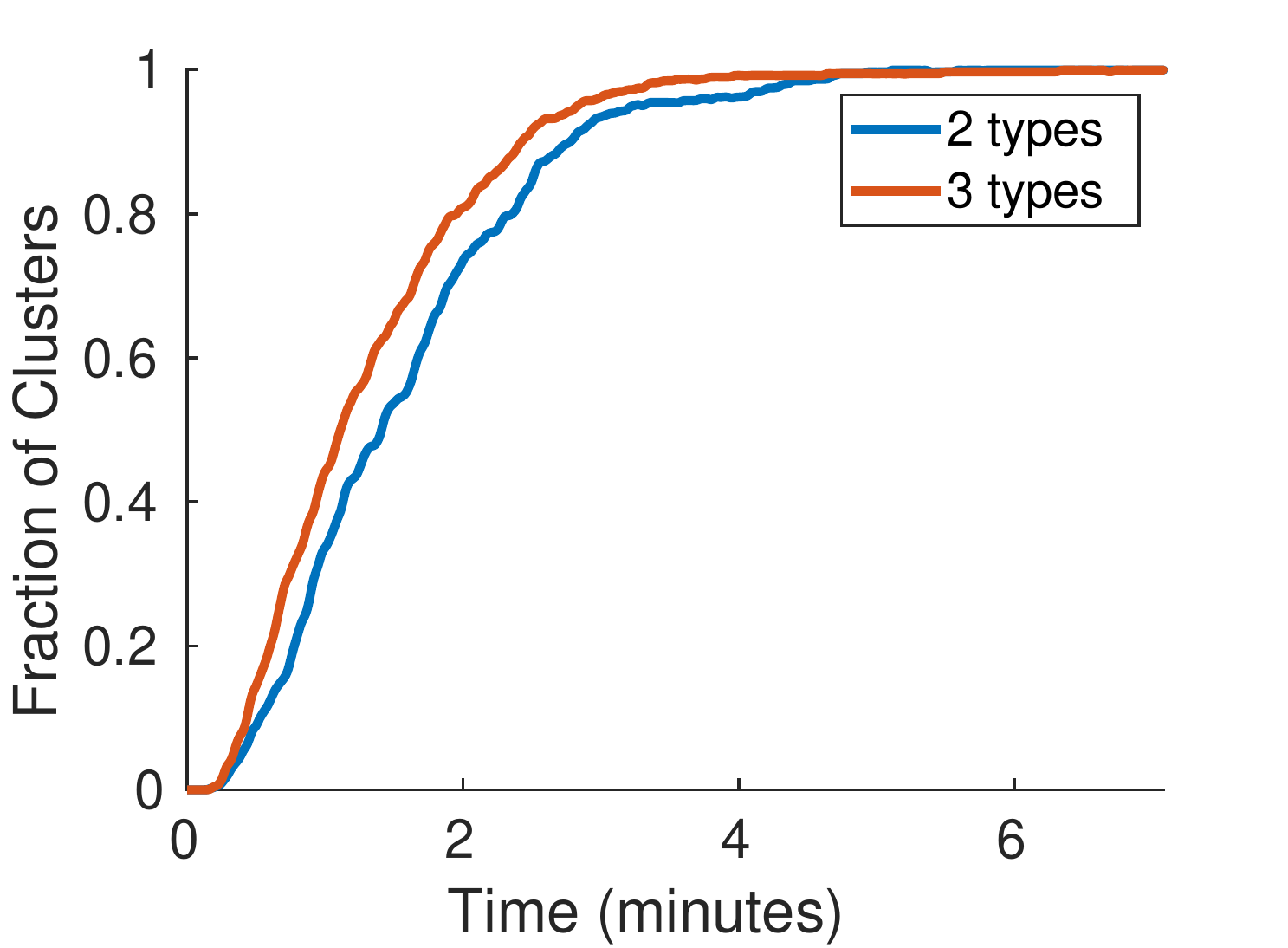}
\label{fig:chevSims}
}
\subfigure[]{\includegraphics[width=0.45\linewidth]{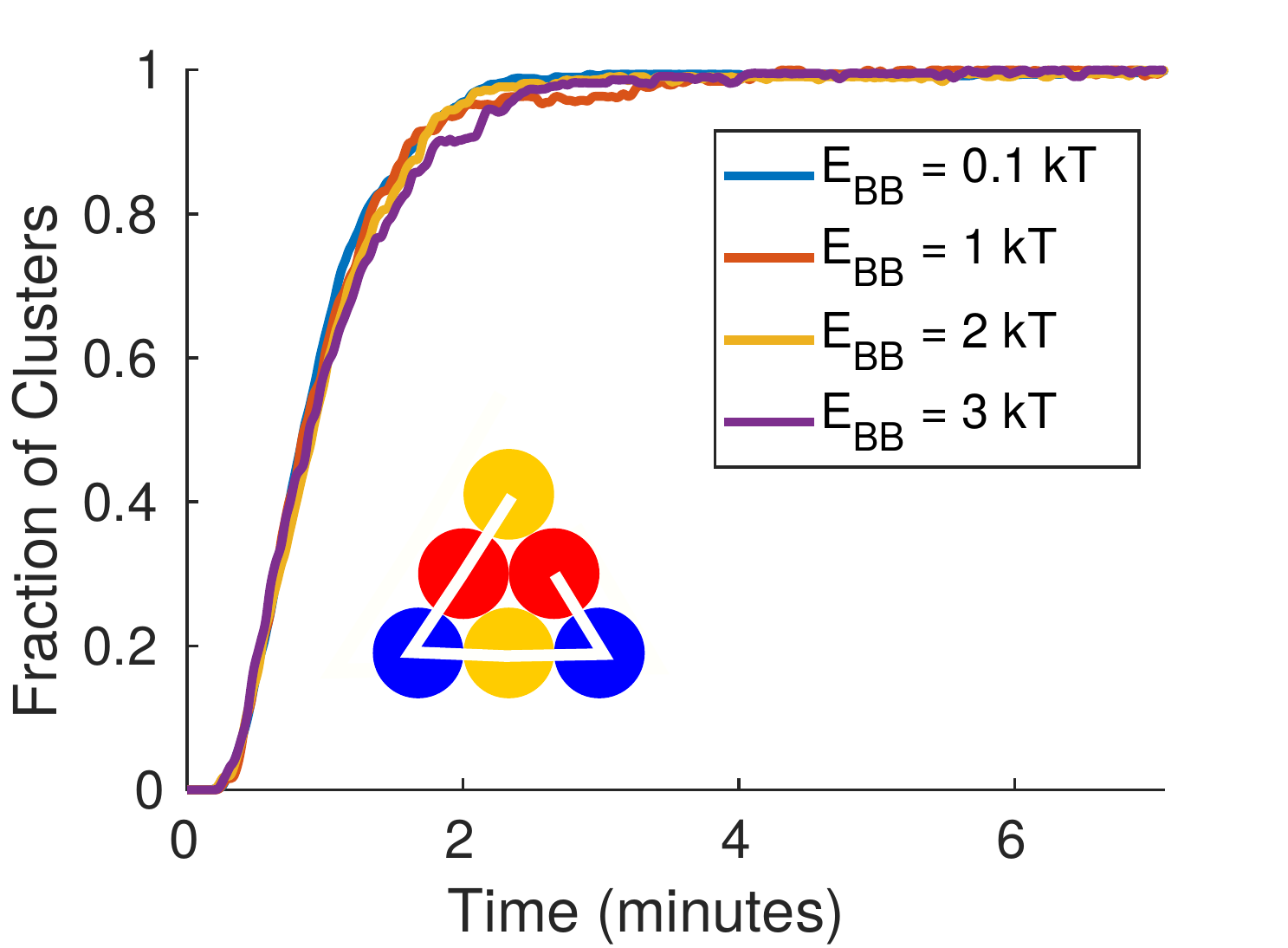}
\label{fig:triangleSims}
}
\caption{(a) Results of a Brownian dynamics simulation of $400$ AAAAAB ($2$ particle types) and AAABCB ($3$ particle types) chains of $6$ colloidal disks, initialized in a linear chain. The interactions were chosen in accordance with the optimal solution for the chevron computed by the genetic algorithm. (b) Results of a Brownian dynamics simulation of $400$ ABCBAC chains of $6$ colloidal disks, initialized in a linear chain. Interaction energies were chosen at various points along the Pareto front, with $E_{AA} = E_{AC} = 12$, and $E_{BB}\in \{0.1, 1, 2, 3\}$, with all others being set to $0.1$. Result is conditioned on ending in the triangle, in order to compare rates. 
The fraction of clusters in each state is plotted as a function of physical time, computed by scaling the non-dimensional simulation time by an appropriate dimensional constant.  }
\end{figure}

We ran Brownian dynamics simulations to test some of the more interesting findings of our coarse-grained model, to see if they are indeed true or an artifact of our model assumptions or numerical algorithms. 

Figure \ref{fig:paretoCollection}(b) shows that a vertical Pareto front is possible with both two and three particle types, but a higher rate is attainable with three types. To test this, we simulated the assembly of the chevron under the optimal interactions found for both two and three particle types.  Figure \ref{fig:chevSims} shows the fraction of clusters in the chevron state as a function of estimated physical time, and we see that there is indeed a slightly higher formation rate with three types. We estimate the mean first passage time to the chevron to be about $1.57$ minutes for two types, and $1.31$ minutes for three types, which is about $17\%$ shorter when using three types. This is likely because the two type chain usually folds by wrapping around the lone B-type particle, proceeding one bond at a time, whereas the three type chain can form several sub-units simultaneously, speeding up the process. 

Another interesting prediction was that the inclusion of some weak, auxiliary bonds not present in the target state could speed up formation of the target. As an example, we found a near vertical Pareto front for the six-disk triangle state using $m=3$ particle types in the configuration ABCBAC. The optimal bond energies were found to be a large $E_{AA}$ and $E_{AC}$, while the rest remain weak. All but one of these weak bonds remained constant along the front, with the one exception being $E_{BB}$, which parameterized the near vertical front, increasing in rate as $E_{BB}$ was increased. We performed Brownian dynamics simulations using the optimal parameters and $E_{BB}\in \{0.1, 1, 2, 3\}$. As stated before, the yield of the triangle is only about $0.5$, so in \ref{fig:triangleSims} we plot the fraction of clusters in the triangle state as a function of time, conditioned on ending up in the triangle state. We see that the value of $E_{BB}$ from our test set has little to no effect on the rate of formation of the triangle, so we are left to conclude that in this case the vertical Pareto front is probably an artifact of our model.

\subsection{Lattice Polymer}\label{sec:latticeDetails}


We consider a two-dimensional, square lattice model, where particles only interact if they are adjacent on the lattice. Each pair of particles, $(i,j)$, in contact contribute an energy $E_{ij}$, depending on the types of the particles and model input parameters. The system is initialized in a linear chain on the $x$-axis, and an MCMC method is used to update the configuration. The proposal moves consist of an end move, rotating a particle at the beginning or end of the chain about its neighbor, and 
a corner move, in which a particle in a corner flips to the opposite corner, if possible. During each MCMC step, all valid moves for the chain are listed, and a proposal is selected uniformly at random from these possibilities. Proposal moves are accepted in accordance with the Metropolis-Hastings acceptance probability. Dynamical time is measured in MCMC steps. 


We again form our coarse-grained model for this lattice system, lumping by adjacency matrix, and use the model to evaluate equilibrium probabilities and mean first passage times. Since these systems are highly degenerate, and often one is interested in a particular permutation, we choose individual permutations as the target state, instead of lumping them into a target set. We study the system with $N=8$ particles on the lattice; the smallest system that can form geometrically distinct ground states. We choose one permutation of each of the two ground states, and apply our genetic algorithm to characterize their Pareto fronts. 

\subsection{Sampling for Larger Systems}\label{sec:Larger}

 Our goal will be to compute the parameterization of the previous Pareto fronts without relying on the coarse-grained model. We do this by taking a sampling approach to evaluate the competing measures for the genetic algorithm. An immediate issue presents itself; equilibrium simulations will take exceedingly long in the presence of kinetic traps, making both of our measures infeasible to compute in a reasonable amount of time. 

One way to make the sampling computationally tractable is to introduce different functions for the objectives. If we can find functions that are correlated with the objectives, preserve the structure of the Pareto front, and are efficient to sample, then we can extract the optimal transformed parameterizations. There are many such functions that can be used, but here we consider just two that have shown promising results. 

We replace $\pi_S$, the equilibrium probability of the target set, by a probability related to staying in the target state once it has formed, $p_s$. Each bond in the target state has an energy, $E_i$, for $i = 1,\cdots,b$, where $b$ is the number of bonds. We compute the harmonic sum of these energies, $h(\vec{E}) = \left(\sum_{i=1}^b E_i^{-1}\right)^{-1}$, which gives a notion of an average energy barrier out of the target state. We then say the probability to stay in the target state is $p_s(\vec{E}) = 1-\exp(-h(\vec{E}))$. Note that this quantity does not have to be sampled, it can be directly evaluated given the target state and input parameters. If all the bonds are weak, $p_s\rightarrow 0$, and if all the bonds are strong, $p_s\rightarrow 1$. 

We replace our rate, $\tau^{-1}$, with a measure related to the kinetic accessibility of a given target state, called $k_A$. To do so, we define two times, $t_{\text{trap}}$ is the maximum amount of time to spend in one state until the system is considered ``trapped'', and $T > t_{\text{trap}}$ is the total simulation time. During a simulation, if the system spends longer than $t_{\text{trap}}$ in a state, without breaking or forming any additional bonds, we stop and compute the \emph{misfolded energy}, $E_{\text{misfold}}$. That is, we add up the energy of all bonds not consistent with the target state using the trapped configuration. If no trap states form, we set $E_{\text{misfold}} = 0$. We average this energy over many trajectories to compute $k_A = \exp(-\langle E_{\text{misfold}}\rangle)$. With this measure, if the system is able to easily escape from misfolded states, $k_A\rightarrow 1$, and if it gets stuck in deep kinetic traps, $k_A\rightarrow 0$. 

Using the coarse-grained Markov model, we are able to evaluate these measures semi-analytically. There is a slight ambiguity in how we define a ``trapped'' state. Given a set of bond energies, $\vec{E}$, we construct the corresponding rate matrix and sort the diagonal entries in absolute value from least to greatest, which give the rates of exiting each state. Typically there is a large gap in this sorted list, so we consider all states before this gap ``trapped'', on a case-by-case basis. We can then evaluate our rate measure by taking a dot product of the misfolded energy vector with a hitting probability vector, which can be computed by solving a matrix equation similar to the linear system we solve for $\tau$.

We revisit the $N=8$ particle, $2$-type, rectangular lattice protein from Figure \ref{fig:paretoLattice}(a), and study its behavior under the new measures. Figure \ref{fig:samplingN8} shows the Pareto front in the original measures (solid blue) and how it maps to the new measures (solid red). We see a large region where $k_A = 1$, which is due to a lack of kinetic traps, according to our above definition. There is a sharp transition to a region in which $k_A$ continuously decreases as $p_S$ increases, meaning the Pareto front is preserved. We then apply our genetic algorithm to the new measures, where we estimate $k_A$ by sampling, using the optimal particle ordering AABABABB. We use a population of size $100$, a limit of $200$ generations, final time $T=500$, a trap time $t_{\text{trap}} = 300$, a mating cutoff of the top $p=25\%$, and $500$ samples to estimate $k_A$ for each population member. We do not observe convergence of the whole population, for reasons we will discuss, so we report only the non-dominated portion of the population, shown as unfilled red circles. 

The sampled Pareto front is close to the analytically computed curve, but seems to give overestimates across most of the range. This is because of how we create the next generation. A non-dominated member of generation $n$ is carried to generation $n+1$, without re-estimating its objective values. This has the effect of only keeping the maximum estimates of $k_A$ for a given range of $p_S$ values, since values of $k_A$ closer to the true average will be dominated by the larger estimates. 

Despite the over-estimate of the position of the actual Pareto fronts, the parameterizations of the front are unaffected by the statistical bias towards larger $k_A$. We extracted the parameters for each of the sampled points along the $(p_S,k_A)$ Pareto front and used the coarse grained model to evaluate the corresponding values of $(\pi_S,\tau^{-1})$, which are plotted as blue, unfilled circles. We find that these points mostly lay on the original Pareto front, confirming that we find the same parameterization. Some of the points lay on the $\pi_S$ axis, with vanishingly small values of $\tau^{-1}$. This is an issue with our choice of $p_S$; we can maximize $p_S$ by making each of the bonds energies as large as possible, but doing so reduces $\pi_S$ because of entropic considerations. 
We handle this issue by discarding
population members that are trying to maximize all bond energies. 

The above tests were performed using the already known optimal particle ordering. We also performed tests where the particle types are also selected by the genetic algorithm. In this case, the performance was significantly worse; we see no convergence and only a handful of points become non-dominated. We do note however, that the non-dominated points are the ones that find the optimal particle ordering. Based on this observation, we propose a two step procedure. An initial run can be used to determine possible candidates for the optimal ordering, and a secondary run can use these candidate orderings to search for a Pareto front. 

As a final test, we apply the sampling approach directly to a larger problem in which enumeration would be expensive. For an $N=16$ lattice protein, we set the target state to be a square and choose the permutation in which the backbone zig-zags up four units and then down four units repeatedly. An optimal configuration using $m=8$ particle types is easy to identify by hand; group particles such that every two have the same type, i.e. AABBCCDDEEFFGGHH. Using the genetic algorithm, the minimum number of types required to achieve a value $k_A > 10^{-3}$ seems to be $m=5$, with the configuration CDDBABBEBDABCBBD. We can again identify Pareto fronts, but they are less informative for this example; the presence of chiral traps begins to skew our $k_A$ measure for this larger system. There are so many chiral traps in this case that even if distinct particles are used, more than $90\%$ of trajectories will get stuck in a state that is not the target. To test whether the Pareto optimal parameters we've computed actually result in efficient assembly, we estimate the yield of the target states compared to the most common kinetic traps for each of the lattice structures we have studied. Figure \ref{fig:latticeTraps} shows this yield comparison, using the parameters found by the genetic algorithm that give assembly comparable to distinct particles. We see that the presence of chiral traps result in a non-negligible reduction of the yields for $N=8$, and the effect gets worse as the system size increases, highlighting the importance of specifying bond angles even more.

\end{document}